%% file: main.tex
\documentclass[10pt,journal,compsoc,onecolumn]{IEEEtran}

\input{our-contents/use_packages.tex}
\input{our-contents/macros.tex}

\ifCLASSINFOpdf
\else
\fi
\begin{document}
%
\title{Workflow Provenance in the Lifecycle of Scientific Machine Learning}
%
%
%
%



\author{
Renan Souza\textsuperscript{1,*},
Leonardo G. Azevedo\textsuperscript{1},
V\'{i}tor Louren\c{c}o\textsuperscript{1},
Elton Soares\textsuperscript{1},
Raphael Thiago\textsuperscript{1},\\
Rafael Brand\~{a}o\textsuperscript{1},
Daniel Civitarese\textsuperscript{1},
Emilio Vital~Brazil\textsuperscript{1}, 
Marcio Moreno\textsuperscript{1},\\
Patrick Valduriez\textsuperscript{3},
Marta Mattoso\textsuperscript{2},
Renato Cerqueira\textsuperscript{1},
Marco A. S. Netto\textsuperscript{1}
\newline{} \newline{}
\textsuperscript{1}IBM Research \\ \textsuperscript{2}Federal University of Rio de Janeiro, Brazil \\ \textsuperscript{3}Inria, Univ. Montpellier, CNRS \& LIRMM, France

\thanks{
This is an authors' preprint of a paper published at Concurrency Computation: Practice and Experience. Please cite it as follows:
\newline{}
{\normalfont Souza, R, Azevedo, L. G., Louren\c{c}o, V., et al. Workflow provenance in the lifecycle of scientific machine learning. Concurrency Computat Pract Exper. 2021;e6544. \url{https://doi.org/10.1002/cpe.6544}}
\newline{}
*Correspondence: Renan Souza - contact@renansouza.org
}
}

\IEEEtitleabstractindextext{%
\begin{abstract}
\input{our-contents/sections/abstract.tex}

\end{abstract}

}

\maketitle

\IEEEdisplaynontitleabstractindextext

%
\IEEEpeerreviewmaketitle


\input{our-contents/index.tex}
\ifCLASSOPTIONcaptionsoff
  \newpage
\fi

\end{document}

%% file: our-contents/use_packages.tex
\usepackage[inline]{enumitem}

\usepackage[disable,colorinlistoftodos,prependcaption,textsize=normal]{todonotes}
\usepackage{xargs}

\usepackage[ruled,vlined,linesnumbered,algo2e]{algorithm2e}

\usepackage{float} 

\usepackage{subcaption} 

\usepackage{textcomp}
\usepackage{listings}
\lstdefinestyle{mystyle}{
    commentstyle=\color{gray}\ttfamily,
    numberstyle=\tiny\color{darkgray},
    stringstyle=\color{purple},
    basicstyle=\footnotesize,
    breakatwhitespace=false,         
    breaklines=true,                 
    captionpos=b,                    
    keepspaces=true,                 
    numbers=left,                    
    numbersep=5pt,                  
    showspaces=false,                
    showstringspaces=false,
    showtabs=false,                  
    tabsize=1,
    upquote=true,
    frame=single,
    morekeywords={as, include, printf, a}
}
\usepackage{color, colortbl}
\usepackage{hyperref}
\usepackage{booktabs}

%% file: our-contents/macros.tex
\newcommand{\eg}{\emph{e.g.},}

\newcommand{\ie}{\emph{i.e.},}

\newcommand{\etal}{\textit{et al.}}

\newcommand{\codefont}[1]{\texttt{#1}}

\newcounter{qcounter} 

\newcounter{DDPcounter} 
\newcommand{\createDDP}[2]{\refstepcounter{DDPcounter}\label{#1}\textit{\textbf{D\ref{#1}: #2.}}}

\newcommand{\refDDP}[1]{\textit{D\ref{#1}}}

\newcounter{SDPcounter} 
\newcommand{\createSDP}[2]{\refstepcounter{SDPcounter}\label{#1}\textit{\textbf{S\ref{#1}: #2.}}}

\newcommand{\refSDP}[1]{\textit{S\ref{#1}}}

\newcommand{\MLCycle}{lifecycle of scientific ML}
\newcommand{\MLCycleCammel}{Lifecycle of Scientific ML}

\newcommand{\MLCycleUPPER}{LIFECYCLE OF SCIENTIFIC ML}

\newcommand{\MLView}{\codefont{MLHolView}}

\newcommand{\queries}{\hyperref[tab:queries]{Q1--Q7}}

{

  \begin{definition}{\textbf{#2}.}  \label {#1}
}
{
\end{definition}
}

\newcommand{\algspace}[1][.1pt]{\par\vskip.4\baselineskip \par\vskip.3\baselineskip}

\newcommand{\ourvspace}{\vspace{0em}}

\definecolor{table_header}{gray}{0.8}
\definecolor{table_row_even}{rgb}{1.0, 1.0, 1.0}
\definecolor{table_row_odd}{gray}{0.9}

\newcommand{\topsepremove}{\aboverulesep = 0mm \belowrulesep = 0mm}

\newcommand{\revisedcolor}[1]{{#1}}

%% file: our-contents/sections/abstract.tex
\noindent 
Machine Learning (ML) has already fundamentally changed several businesses. More recently, it has also been profoundly impacting the computational science and engineering domains, like geoscience, climate science, and health science. In these domains, users need to perform comprehensive data analyses combining scientific data and ML models to provide for critical requirements, such as reproducibility, model explainability, and experiment data understanding.  However, scientific ML is multidisciplinary, heterogeneous, and affected by the physical constraints of the domain, making such analyses even more challenging. In this work, we leverage workflow provenance techniques to build a holistic view to support the lifecycle of scientific ML. 
We contribute with (i) characterization of the lifecycle and taxonomy for data analyses; (ii) design decisions to build this view, with a W3C PROV compliant data representation and a reference system architecture; and (iii) lessons learned after an evaluation in an Oil \& Gas case using an HPC cluster with 393 nodes and 946 GPUs.
The experiments show that the decisions enable queries that integrate domain semantics with ML models while keeping low overhead (\textless1\%), high scalability, and an order of magnitude of query acceleration under certain workloads against without our representation.

%% file: our-contents/index.tex
\input{our-contents/sections/introduction}

\input{our-contents/sections/lifecycle}

\input{our-contents/sections/design_principles}

\input{our-contents/sections/prov_data_representation.tex}

\input{our-contents/sections/experiments}

\input{our-contents/sections/related_work}

\input{our-contents/sections/conclusion}

\bibliographystyle{IEEEtran.bst}
\bibliography{our-contents/references.bib}

\input{our-contents/sections/appendices.tex}

%% file: our-contents/sections/introduction.tex
\section{INTRODUCTION} \label{sec:intro}

Machine Learning (ML) has been fundamentally transforming
several industries and businesses in numerous ways. More recently, it has also been impacting computational science and engineering domains, such as geoscience, climate science, material science, and health science.
Scientific ML, \ie{} ML applied to these domains, is characterized by the
combination of data-driven techniques with 
domain-specific data and knowledge
to obtain models of physical phenomena~\cite{MLCSE, gil_intelligent_2018, raissi_pinns_2019, rodrigues2018deepdownscale,salles19}.
Obtaining models in scientific ML works similarly to conducting
traditional large-scale computational experiments \cite{mattoso_scientific_2010},
which involve a team of 
scientists and engineers that 
    formulate hypotheses, 
    design the experiment and predefine parameters and input datasets,
    analyze the experiment data, 
    do observations, and 
    calibrate initial assumptions in a cycle until they are satisfied with the results.
Scientific ML is naturally large-scale because multiple people collaborate in a project, using their multidisciplinary domain-specific knowledge to design and perform
data-intensive tasks to curate (\ie{} understand, clean, enrich with observations) datasets and prepare for learning algorithms. They then plan and execute compute-intensive tasks for computational simulations or training ML models affected by the scientific domain's constraints.
They utilize specialized scientific software tools running either on their desktops, on cloud clusters (\eg{} Docker-based), or large HPC machines.

Other works propose an ML lifecycle \cite{modelhub_icde2017,schelter2017automatically}.
Although they might apply for scientific ML, in our view, 
there are still gaps in these lifecycle proposals to properly address scientific ML characteristics, particularly
the need for deeper integration with scientific domain data and specialized knowledge on a domain.
Our proposed model for the \textit{\MLCycle{}} has three phases (explained in detail later in this paper):
    \textit{data curation} --- to curate raw data;
    \textit{learning data preparation} --- to prepare the curated data for learning;
    and the \textit{learning} itself --- aware of the constraints of a scientific domain.
In each of these phases, there may be multiple workflows. Each workflow is a set of chained data transformations consuming and producing datasets, and a workflow may consume the datasets produced by another workflow.
For instance, there may be multiple workflows only in the learning data preparation phase to transform curated data into learning datasets. These datasets may then be consumed by multiple workflows in the learning phase, transforming the datasets into different ML models.
Therefore, we propose modeling these workflows as multiple interconnected workflows \cite{provlake_escience_2019}.
From now on, we refer to \textit{workflows} as these multiple interconnected workflows in all phases of the \MLCycle{}.

Our primary goal in this paper is to support this lifecycle by enabling scientists and engineers to perform comprehensive, \ie{} end-to-end data analyses that integrate the data consumed and generated in these workflows, from raw domain data to learned models.
The importance of these data analyses is that they are enablers to meet 
critical requirements in ML, such as model reproducibility and explainability, and experiment data understanding.

The main problem to achieve this goal is to deal, in an integrated and comprehensive way, with the high heterogeneity of different contexts 
(\eg{} data, software, environments, persona) involved in this lifecycle. 
For example, the analyses need to be aware of 
the (hyper)parametrization of different data transformations in various workflows, how the transformations affect the experiment results (\eg{} quality of the ML models), and the relationships between parameters, results, and domain-specific data and knowledge. 
For instance, one may ask: ``what happened to the model performance when the parameters varied from X to Y when the datasets had a specific characteristic in the domain?''.  
To allow for such analyses, tracking how the data are transformed throughout the workflows in an integrated and holistic way is necessary. 
Not having such holistic integration is critical for several reasons. To exemplify, it compromises experiment reproducibility from a scientific perspective. From a business perspective, stakeholders may be less likely to apply an ML model, even with the best performance, if they do not understand the transformations that led to the best model)\cite{MLLifeCycle}.

Provenance (also referred to as lineage) data management techniques help reproduce, trace, assess, understand, and explain data, models, and their transformation processes~\cite{herschel_survey_2017,ccpe_first_prov_challenge_2008, buneman2019data}.
The provenance research community has evolved significantly in recent years to provide for several strategic capabilities,
    including experiment reproducibility~\cite{thavasimani2016facilitating}, 
    user steering (\ie{} runtime monitoring, 
    interactive data analysis, runtime fine-tuning)~\cite{souza_keeping_2019}, raw data analysis~\cite{Silva2016AnalyzingRawCCPE}, and our previous work, which helps data integration for multiple workflows generating data in a data lake~\cite{provlake_escience_2019}.
Furthermore, other works contribute to support provenance tracking specifically for ML workflows~\cite{miao_towards_2017, zaharia_accelerating_2018, debora_sbbd_2019,schelter2017automatically,modelhub_icde2017}, including reproducible models and explainability~\cite{lucero2018exploring}.
These related works are essential building blocks to be leveraged towards supporting the lifecycle.

Nevertheless, scientists and engineers
still face difficulties in performing comprehensive data analyses that would help them meet those critical requirements in ML.
Tracking provenance in those workflows could
be used as a tool to provide for a holistic view, hence enabling the data analyses. However, the problem caused by the high heterogeneity in the lifecycle arises several challenges. 
For example, the workflows are highly heterogeneous and with distributed execution control: there may not be one single Workflow Management System (WMS) orchestrating all workflows; instead, there may be multiple WMSs, scripts, programs, and ML and data processing frameworks without a single unified execution orchestrator. 
Further, these workflows manage domain- and ML-specific data and knowledge stored in various distributed data stores and run on various execution environments. Hence, strategies to track data in multiple data stores are needed.
Also, another complicating factor is that efficiency is a common requirement, especially in HPC executions. Thus the systems supporting the lifecycle need to scale and not add significant tracking overhead.
Designing a system to efficiently track provenance in such heterogeneous scenarios has been recently acknowledged as a research challenge by leading data management researchers~\cite{sigmodpanel_2020}.

In this paper, our focus is to support the \MLCycle{} by proposing an approach for comprehensive data analyses, addressing the problem of high heterogeneity of different contexts. Particularly, we contribute with: 

\begin{enumerate}[label=(\roman*),nolistsep]
    \item A comprehensive characterization of the lifecycle, from raw domain data to learned models passing through the processes that manipulate these data, and a taxonomy (detailing \eg{} data, execution timing, and training timing classes) positioning the role of provenance analysis to support the lifecycle  (Sec.~\ref{sec:ml_lifecycle});
    
    \item Data design decisions to build and query a provenance-based holistic data view to integrate data, processed by workflows in the lifecycle, aware of the heterogeneous dimensions and enable comprehensive analyses; and architecture decisions that guide how to build a provenance system to efficiently track and integrate data in distributed executions (Sec.~\ref{sec:principles}); 

    \item PROV-ML, a new provenance data representation for scientific ML leveraging W3C PROV\cite{W3CPROV} and MLS\cite{W3CML} (Sec.~\ref{sec:provml});
    
    \item Lessons learned after applying the design decisions in a system's implementation and evaluating it in a real case in the Oil \& Gas (O\&G) industry in a testbed with 3 environments, including an HPC cluster with 393 computing nodes and 946 GPUs. 
    We found that the decisions enabled comprehensive queries with rich semantics about the application domain and ML while maintaining low tracking overhead (\textless1\%), near-linear scalability, and efficient query performance (over an order of magnitude compared with a provenance representation without PROV-ML for certain workloads) (Sec.~\ref{sec:exps})
    
\end{enumerate}

\noindent
Finally, we present the related work in Section~\ref{sec:related_work}, and conclude in Section~\ref{sec:conclusion}.

\revisedcolor{
The major extensions related to our work published on IEEE WORKS@SC19~\cite{souza_provml_2019} are as follows: 
(i) We detail the explanations of the decisions that drive the data and architectural design of a system to support data analyses in the lifecycle. These details better highlight the lessons learned so that other researchers and practitioners can learn from them;
(ii) We refine and extend PROV-ML to better represent subparts of the learning process and to make the model clearer by using more accurate names for the concepts, refining  descriptions, and changing a few structures of the representation;
(iii) We include a new set of experiments, which demonstrate that PROV-ML can yield query acceleration allowing better experience while using the system for data analyses; and (iv) We discuss how the proposed approach can be customized to different applications and domains.
}

%% file: our-contents/sections/lifecycle.tex
\ourvspace{}
\section{CHARACTERIZING THE \MLCycleUPPER{}}
\label{sec:ml_lifecycle}

Existing works describe an ML lifecycle \cite{modelhub_icde2017,schelter2017automatically}, but such descriptions focus on business domains and do not address the high heterogeneity problem of the  \MLCycle{}.
Since our main goal is to support this lifecycle by enabling scientists and engineers to perform comprehensive data analyses, we begin with a proposal of a model for this lifecycle and a thorough characterization.
This is the first work that proposes a lifecycle focused on scientific ML to the best of our knowledge. 
We first characterize the personas and describe the lifecycle  (Sec. \ref{subsec:mllifecycle}), then we present our motivating use case (Sec.  \ref{subsec:usecase_description}), and finally, we characterize the data analyses using provenance (Sec. \ref{subsec:provanalysis}).

\ourvspace{}
\subsection{The \MLCycleCammel{}} 
\label{subsec:mllifecycle}

\begin{figure}[ht]
  \centering
  \includegraphics[width=0.6\linewidth]{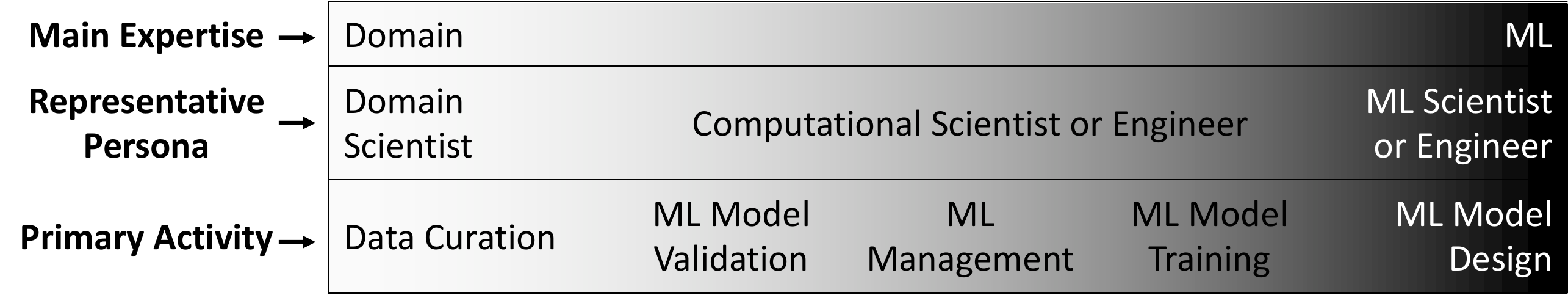}
  \caption{Spectrum of expertise and personas in the lifecycle.}
   \label{fig:personas}
   \ourvspace{} 
\end{figure}

Multidisciplinary personas, with different skills in the domain and ML techniques, participate in the lifecycle phases.
In our previous work, we presented a spectrum of expertise and personas in scientific ML~\cite{souza_provml_2019}, depicted in Figure~\ref{fig:personas} and briefly summarized here.
The spectrum ranges from only scientific-domain (fully white on the left) to ML only (fully black on the right), with the following personas:
\begin{enumerate*}[label=(\roman*)]
    \item \textit{Domain scientists}, who have in-depth knowledge of the domain data and use specialized tools to interpret, visualize, and clean the scientific data;
    \item \textit{Computational scientists and engineers}, who have high computational skills, often with abilities to develop parallel scripts and execute them in HPC clusters; \item \textit{ML scientists and engineers}, who have in-depth knowledge of statistics, ML algorithms, and software engineering. 
\end{enumerate*}
In an orthogonal sense, \textit{Provenance specialists} design the provenance schema for applications and guide other users to add provenance capture hooks to the workflows. 

\begin{figure}[ht]
  \centering
  \includegraphics[width=0.8\columnwidth]{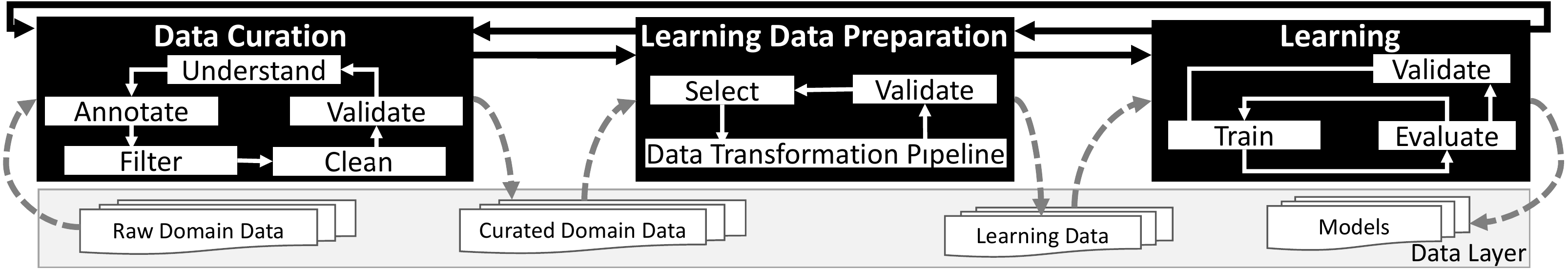}
  \caption{The \MLCycleCammel{}.}
   \label{fig:lifecycle}
   \ourvspace{}
\end{figure}

Our proposed model of the \MLCycle{} is to divide it into three phases: \textit{data curation}, \textit{learning data preparation}, and \textit{learning} (Figure~\ref{fig:lifecycle} --- dashed arrows are data flows and solid arrows are interactions between phases). 

\smallskip
\noindent \textbf{Data curation.} 
It is the most complex phase of the lifecycle, mainly because of the nature of the scientific data. 
Much manual and highly specialized work are performed by the users (primarily domain scientists) to achieve automated knowledge extraction from scientific data promoted by ML. 
There is a significant gap between raw scientific data and useful data for consumption (\eg{} data to serve as input to train ML models). 
Datasets can be huge, typically containing geospatial-temporal data stored in scientific formats, like HDF5, NetCDF, SEG-Y. 
Specialized formats in scientific domains may require industry-specific software and domain-specific knowledge to inspect, visualize, and understand the data. 
In addition, users can use metadata and textual reports to annotate the data with extra domain-specific knowledge, without which would be nearly impossible to make the data useful for ML algorithms. 
Considering the heterogeneous nature of the data, ``it is unreasonable to assume that data lives in a single source'' (\eg{} a single file system or DBMS) \cite{MLLifeCycle}. 
For instance, raw files can be stored in file systems or cloud stores, domain-specific annotations can be stored in a Knowledge Base System (\eg{} Triple Store) with domain ontologies, and curated data can be stored in a NoSQL DBMS. 
Then, computational scientists and engineers develop data-intensive scripts to clean, filter, and validate the data. 
Each of these steps inside the data curation phase is highly interactive, manual, and may execute independently.
In other words, users may run different scripts to perform these phases, several times, in an \textit{ad-hoc} way, in any order, and on different machines.
These phases occur in a cycle, which stops when the users consider the data ``curated''. 
In the context of ML, it is ready to be transformed into learning data.


\smallskip
\noindent \textbf{Learning data preparation.} Model trainers select relevant parts of the curated data to be used for learning. 
For instance, if the ML task is to classify geological structures~\cite{salles19}, seismic images will need to be correlated with seismic interpretation, creating labeled samples. 
After selecting the data, model designers develop scripts,
typically using domain-specific libraries to manipulate the raw scientific data,
to transform (\eg image cropping, quantization, scaling) the data into learning datasets. 
Due to data complexity, frequently, data need to be manually inspected before it can be used as input for the learning phase.

\smallskip
\noindent \textbf{Learning.} 
The learning contemplates training, validation, and evaluation.
In this phase, model trainers select the input learning datasets, optionally they choose validation datasets, and choose learning parameters (\eg{} in deep learning they can choose ranges of epochs and learning rates) that will be optimized.
Trainers can use their domain knowledge to discard learning datasets that will unlikely provide good results. The learning process is compute-intensive, typically executed in an HPC machine. 
One single learning process often generates multiple learned models, among which one is chosen as the ``best'' depending on evaluation metrics (\eg{} MSE, accuracy, or any other user-defined metric). 
Moreover, trainers need to monitor the learning process by, \eg{} inspecting how the evaluation metrics are evolving while the learning process iterates. 
They can wait until completion or interrupt the learning process, change parameters, iteratively re-submit the learning until satisfied with results. 

\ourvspace{}

\subsection{\revisedcolor{Motivating Use Case: The Lifecycle of a Deep Learning Classifier in Geoscience}}
\label{subsec:usecase_description}

\revisedcolor{
We explore a motivating use case in the O\&G industry to illustrate the classes of data analyses driven by data integration via provenance capture. 
Finding new reservoirs is a demanding task in the O\&G industry and involves a broad spectrum of actions, such as the interpretation of seismic surveys. 
These surveys are indirect measures of the earth subsurface that can be organized into slices (images). 
They cover hundreds of square kilometers and help to interpret the geology by identifying geological structures, like salt bodies, and find possible hydrocarbons accumulations. 
Processing seismic data imposes complex chained data transformations and can suffer from many problems, like noise and shadows (regions with low signal). 
Trying to automate such activity is of high interest in academia and industry and deep learning is a promising machine learning technique for this \cite{salles19}.
However, the geological structures vary geographically, from point to point in the subsurface, imposing significant challenges to the ML algorithms.
Thus, it requires specialized knowledge to prepare, clean, and understand the data processed in the workflows.

To cope with this, often different teams in an interdisciplinary group composed of geoscientists, computational scientists, engineers, statisticians, and others decompose the problem into parts so that each can address different facets of the problem. Nonetheless, each team has a preferred way to automate tasks and store data, and a team consumes data generated by another.
Despite decomposing the problem into parts makes the problem feasible, it creates a new problem: 
how to analyze the data in an integrated way
\cite{provlake_escience_2019, souza_provml_2019, salles19}. 
Therefore, a unified view over those multiple parts is required so users with various personas can use it to analyze the data. Table \ref{tab:queries} shows seven exemplary data analyses (via queries) that integrate the phases of the lifecycle.

\input{our-contents/tables/queries_table.tex}

Although this use case is in the O\&G industry, we observe a similar demand in several other domains. For instance, designing ML algorithms to handle problems in other industries, such as health, high-energy physics, bioinformatics, and manufacturing~\cite{MLCSE, gil_intelligent_2018, raissi_pinns_2019, rodrigues2018deepdownscale,salles19}. In these areas, the development of ML models typically requires complex computational experiments designed as multiple workflows, executed on different HPC clusters, and also involve collaboration among various experts. Therefore, being able to analyze the experiment data associated with the ML models through all lifecycle phases is essential.
}

\ourvspace{}
\subsection{A Taxonomy of Data Analyses in the \MLCycleCammel{} using Provenance} 
\label{subsec:provanalysis}

Provenance data in workflows contain a structured record of the data derivation paths within chained data transformations and their parameterizations \cite{Silva2016AnalyzingRawCCPE, souza_keeping_2019}. 
Provenance data are usually represented as a directed graph where vertices are instances of entities (data) or activities (the data transformations) or agents (\eg{} users); and, edges are instances of relationships between vertices \cite{W3CPROV}. 
Comprehensive data analysis using provenance has been used as an enabler for several key capabilities:

\begin{itemize}
    \item Experiment reproducibility \cite{herschel_survey_2017, freire2008, ccpe_first_prov_challenge_2008}
    \item AI explainability \cite{lucero2018exploring, miao_sigmod2019, provlake_escience_2019};
    \item Experiment fine-tuning and what-if analyses \cite{souza_keeping_2019};
    \item Uncertainty quantification~\cite{da2017characterization, guerra2012uncertainty};
    \item Hypothesis testing~\cite{mattoso_scientific_2010}; and
    \item Real-time monitoring, and interactive data analysis.~\cite{souza2017data}
\end{itemize}

\noindent 

\begin{figure}[ht]
  \ourvspace{}
  \centering
  \includegraphics[width=\textwidth]{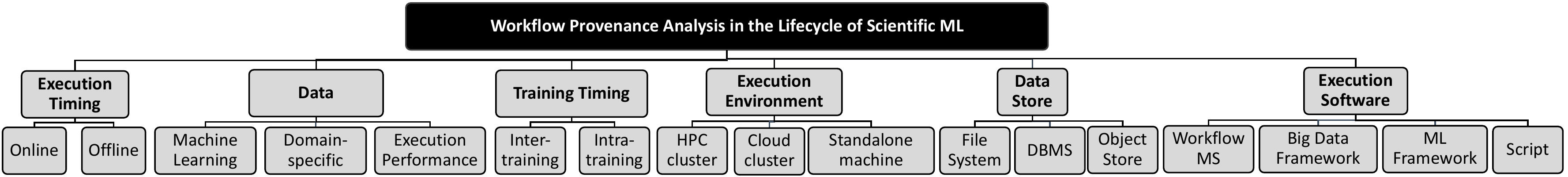}
  \caption{A taxonomy for workflow provenance analysis of the \MLCycle{}.}
  \label{fig:querytaxonomy}
  \ourvspace{}
\end{figure}

\noindent 
Based on a literature analysis~\cite{joao_survey_2019, herschel_survey_2017, MLLifeCycle, sikos_provenance_aware_2020, freire2008} and on our own experience to leverage provenance to support workflows for scientific ML~\cite{provlake_escience_2019, souza_provml_2019,thiago2020managing,souza_aapg_2020}, we propose here a taxonomy (Fig. \ref{fig:querytaxonomy}) to classify workflow provenance analysis in support of ML, by considering three classes: \textit{data}, \textit{execution timing}, and \textit{training timing}. 
Next, we characterize the data involved in the lifecycle.

\noindent \textbf{Data} class includes \textit{domain-specific}, \textit{machine learning}, and \textit{execution}
Provenance data may be augmented with these data, increasing the scope of the analysis.

\textit{Domain-specific data} are the main data processed in the data curation phase (Sec. \ref{subsec:mllifecycle}).
Approaches to add domain data into provenance analysis include, \eg{} raw data extraction \cite{silva_raw_2017} and utilization of domain-specific knowledge databases associated to provenance databases \cite{provlake_escience_2019}. 
For raw data extraction, quantities of interest are extracted from raw data files. 
For domain databases, domain scientists may provide relevant information and metadata about the raw data and store them in knowledge graphs.


\textit{Machine learning data} include learning data and generated learned models, which are more related to the learning data preparation (\eg{} Q1) and learning (\eg{} Q2, Q3, Q7) phases (Fig.~\ref{fig:lifecycle}). These queries exemplify that the parametrization within the data transformations and relevant metadata of the generated data is important for provenance analysis. 




\textit{Execution data.} Besides model performance metrics (\eg{} accuracy), users need to assess workflow execution time and resource consumption. They need to inspect if a critical block in their workflow (\eg{} one demanding high parallelism) is taking longer than usual or if other parts are consuming more memory than expected. For this, provenance systems can capture system performance metrics and timestamps (\eg{} Q4). Metadata, such as data store metadata (\eg{} host address), HPC cluster name, and nodes in use, can be captured and associated with the provenance of the data transformations for extended analysis.


\textit{Hybrid.} These data can be combined. In Q5 and Q7, the analysis queries data processed in workflows in the learning data preparation and learning phases, whereas Q6 uses the same dataset to analyze the raw files curated in the data curation phase.

\smallskip
\noindent \textbf{Execution timing} refers to if the analysis is done \textit{online}, \ie{} while at least a workflow is running, or \textit{offline}. 

\textit{Offline analysis.} The typical use of offline provenance analysis is to support reproducibility and historical data understanding, \eg{} understand the curation of raw files and relate with the ML models. 
For example, the queries \queries{} can be executed offline.

\textit{Online analysis.} Users can use online provenance analysis to monitor, debug or inspect the data transformations while they are still running (\eg{} see the status, see how the intermediate results are evolving as the input parameters vary). 
The problem of adding low provenance data capture overhead is more challenging for provenance systems that allow for online analysis \cite{provlake_escience_2019}. 
Queries Q3--Q5 and Q7 exemplify queries that can be executed online, \eg{} while a training process is running.

\smallskip
\noindent \textbf{Training timing} refers to whether the analysis performs
\textit{intra-training}---\ie{} to inspect one training process, \eg{} a training job running on an HPC cluster, or 
\textit{inter-training}---\ie{} analyses comprehending results of several training processes. 

\textit{Intra-training}. 
In an offline intra-training analysis, users are interested in understanding how well-trained models generated in a given training process perform. The queries \queries{} can be executed either online or offline, but Q3 and Q4 are more likely to be performed as online intra-training analysis.

\textit{Inter-training}. This analysis refers to comprehensive queries to understand multiple training processes, \eg{} 
how each of them performed, 
which learning datasets were used, 
how the training processes were parameterized. 
It supports activities like Model Validation, Management, Training, and Design. 
Usually, they are used offline, but may also be performed online.
Queries \queries{} fit this class when analyzing multiple trained models generated in different training processes.

\noindent \textbf{Further characterization.} Other classes worth mentioning for provenance analysis are: \textit{data store}---data are distributed onto multiple stores, like file systems, cloud stores (\eg{} IBM Cloud Object Storage, AWS S3), Relational or NoSQL DBMSs \cite{provlake_escience_2019}; 
\textit{execution environment}---where the workflows execute, such as HPC clusters, Kubernetes clusters, Standalone server;
\textit{execution orchestration software}---each workflow may be executed as a standalone script, or as a workflow in a WMS, or a composition of microservice calls, or as a pipeline in a data processing (\eg{} Spark) and ML frameworks (\eg{} Tensorflow);
\textit{provenance data granularity}---provenance of files (\ie{} references to files consumed and generated in a script), functions calls (arguments and outputs), blocks of code, and stack traces \cite{joao_survey_2019}; and 
\textit{provenance analysis direction}---forward or backward: generally, forward queries analyze from raw scientific files or learning datasets to trained models (\eg{} Q3--Q5, Q7), whereas backward queries analyze from trained models to learning datasets or raw files (\eg{} Q1, Q2, Q6).

%% file: our-contents/tables/queries_table.tex
\begin{table}[ht] 
\caption{Examples of provenance queries in the \MLCycle{}.}
\footnotesize
\topsepremove
\begin{tabular}{>{\arraybackslash}m{0.01\textwidth}m{0.93\textwidth}}
\rowcolor{table_header}
\textbf{{ID}} & \textbf{Description}
\\
\toprule
\rowcolor{table_row_even}
{Q1} & Given a trained model, what are the geographic coordinates, oil basin and field, and the number of seismic slices of the seismic in the training dataset?
\\ 
\rowcolor{table_row_odd}
{Q2} & Given a trained model, what is the tile size, the noise filter threshold, and the ranges of seismic slices that were selected to generate the training set used to adjust this model?
\\ 
\rowcolor{table_row_even}
{Q3} & Given a training set, what are the values for all hyperparameters and the evaluation measure values associated with the trained model with least loss?
\\ 
\rowcolor{table_row_odd}
{Q4} & What are the average, min, and max execution times of each batch iteration inside each epoch of the deep neural network training, given a training dataset? 
\\ 
\rowcolor{table_row_even}
{Q5} & What is the execution time on average per batch iteration, per epoch, and what are the evaluation metrics of the trained models that used the training dataset generated for a given range of seismic slices? 
\\ 
\rowcolor{table_row_odd}
{Q6} & Given the training dataset used in Q5, what was the seismic data file used, along with its number of slices, related oil basin, and field?     
\\ 
\rowcolor{table_row_even}
{Q7} & 
Considering only the learning workflows that used the learning dataset associated to a given range of seismic slices, list the minimum batch loss per model obtained in the learning stage, also listing the model's hyperparameters and evaluation measurements jointly with the hyperparameters and measurements for associated model obtained in the validation stage, ordered by the best learned models.
\\
\arrayrulecolor{table_row_odd}\hline
\end{tabular}
\label{tab:queries}
\ourvspace{}
\end{table}

%% file: our-contents/sections/design_principles.tex
\ourvspace{}
\section{PROVENANCE IN THE \MLCycleUPPER{}} 
\label{sec:principles}

\revisedcolor{
This section presents the fundamental design decisions for effective and efficient management of workflow provenance data in the \MLCycle{} to provide for comprehensive data analyses.
Although some of these decisions, individually, have been presented in related works~\cite{goble_fair_2019, provlake_escience_2019, hu2019efficient, silva_raw_2017},
together they compose the building blocks of our approach and we assemble them as one unified set and describe how they support the lifecycle.
They are organized as:
\begin{enumerate*}[label=(\roman*)]
    \item \textit{Data Design} (Sec. \ref{subsec:data_designprinciples}), which contains the decisions and key concepts that drive the contents of our holistic data view, whose a resulting artifact is PROV-ML, a new provenance data representation; and,
    \item \textit{System Design} (Sec. \ref{sec:system-design-principles}), which contain the decisions that determine how the provenance data are captured in a scalable and portable manner, whose resulting artifact is a reference system architecture.
\end{enumerate*}
}

\ourvspace{}
\subsection{\revisedcolor{Data Design}}
\label{subsec:data_designprinciples}

\noindent
\createDDP{dp:mlview}{Data Integration with a Holistic Data View}
The primary design decision is that to be able to manage effectively (\ie{} capture, integrate, store, and query) provenance data in the interconnected workflows in all lifecycle phases, a provenance system must implement techniques to provide for an integrated, unified, and holistic data view.
Also, it has to be aware of the contexts of the data transformations in the multiple workflows that consume and generate these data, their (hyper)parameterization, and output values, where these transformations run, where the generated data are stored, who are the involved personas, and how they interact with the workflows.
This design decision builds on a multi workflow data view concept proposed in our previous work \cite{provlake_escience_2019}. It extends it to support the lifecycle comprehensively, with specializations to address ML-specific data and knowledge related to domain-specific data and knowledge.
Let us call this data view as the \textit{Provenance-based Holistic Data View of the \MLCycleCammel{}} (\textbf{\MLView{}}).
The contents and the granularity of the \MLView{} are driven by the relevant queries for a project, and the view can be materialized as the database that integrates data from several sources, while the workflows run \cite{provlake_escience_2019}. 

\noindent
\createDDP{dp:context-awareness}{Context-awareness using Knowledge Graphs: Domain, ML, and Hybrid environments and Data stores}
Extending provenance with domain-specific data for data analysis has been explored before~\cite{oliveira2015,silva_raw_2017,souza_keeping_2019}.
However, in scientific ML, it is required to go a step further into the details of domain-specific knowledge, including how key domain concepts relate to each other. 
Thus, it is important to relate the data in the workflows with as much knowledge as possible available about the project's key concepts.
To be able to integrate with domain-specific knowledge databases, one needs to design the workflows aware that files (or data in other data stores, like DBMSs or object stores) are associated with concepts defined elsewhere. Then the provenance system needs to provide the proper links between the file and the domain-specific concepts. 

Similarly, the \MLView{} needs ML-specific concepts and relationships. 
Although modeling ML-specific concepts could be seen as modeling data for a specific domain (in this case, ML would be the domain),
ML, by itself, is a distinguished domain, which crosses many industries and scientific domains. Thus, the \MLView{} should have a built-in ML-specific schema, tightly coupled with the rest of the provenance data schema, to provide ML-specific context to support the comprehensive analyses. 
In certain cases, such specialized schema modeling might be even helpful to accelerate queries that require them~\cite{lei2020property}. 

\begin{figure}[ht]
  \centering
  \includegraphics[width=0.9\columnwidth]{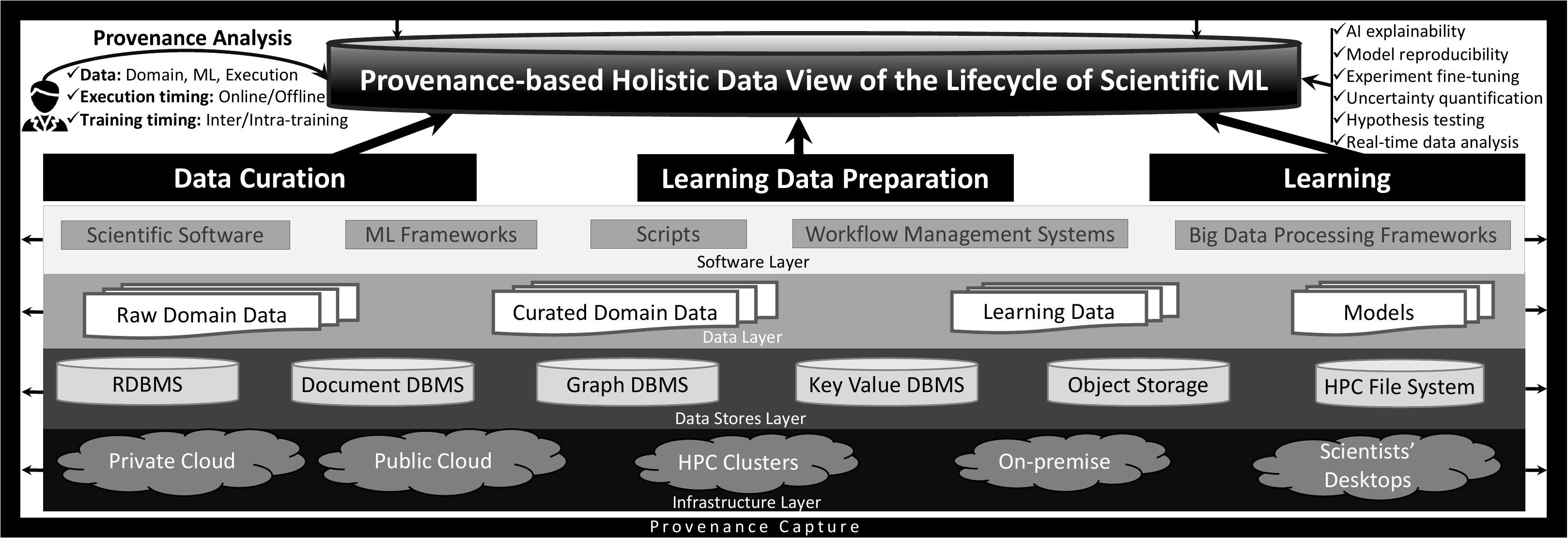}
  \caption{The Provenance-based Holistic Data View of the \MLCycleCammel{}.}
   \label{fig:mlholview}
\end{figure}

In addition to the domain- and ML-specific context awareness, 
since the workflows can be executed within heterogeneous frameworks, scripts, or WMSs and on heterogeneous environments, 
the \MLView{}  needs to be aware of such hybrid (\ie{} heterogeneous) execution by containing the track of the execution environment and software, and associated metadata.
These data and their relationships, with pointers to domain-specific knowledge graphs and large data stored in other stores, are all materialized using provenance data in the knowledge graph that forms the \MLView{}.
Figure~\ref{fig:mlholview} illustrates the \MLView{} and its awareness of data coming from the ML phases and the dimensions of heterogeneity (illustrated as layers) it addresses: software, data, data stores, and infrastructure (execution environment). 
The figure also shows the kind of provenance analysis (\textit{top-left}) and the key capabilities the \MLView{} enables (\textit{top-right}) (Sec. \ref{subsec:provanalysis}). 

\noindent
\createDDP{dp:provschema}{Provenance of Multiple Workflows on Data Lakes meets ML Provenance Following W3C Standards}
To be able to implement the context-awareness for the domain, ML, and hybrid environments
, the \MLView{} needs a comprehensive data representation.
Data lake provenance builds on workflow provenance to enable the awareness of the location of each data item generated by chained data transformations in a data lake, even if there are multiple data items dispersed in hybrid environments and data stores \cite{provlake_escience_2019, komadu_escience2016}, making it a good alternative to address such heterogeneity of data, store, and environments.
However, it is not enough to support the lifecycle, as it requires provenance of ML-specific data and learning processes.

The provenance data community has significantly evolved in recent years, oftentimes leveraging the PROV \cite{W3CPROV} family of documents, a W3C recommendation, making it a  \textit{de facto} standard that provides the building blocks, in terms of data representation, for any provenance-based approach, allowing for compatibility among different solutions~\cite{missier2010linking}.
The PROV-Wf~\cite{costa2013capturing} workflow provenance data representation and its derivatives~\cite{Silva2016AnalyzingRawCCPE} have also been used and evolved by several initiatives~\cite{souza2017data, souza_keeping_2019, souza2018provenance, souza_distributed_2021}.
Our previous work builds on W3C PROV and PROV-Wf to propose PROVLake, a first provenance data representation for workflows on data lakes \cite{provlake_escience_2019}. 
Concerning ML-specific data modeling, there is a W3C community group developing a data representation with specific ML vocabulary, the W3C ML~Schema~(MLS)~\cite{W3CML}. 
Therefore, this data design decision proposes that
the data representation for the \MLView{} should be comprehensive,
with detailed semantics about the workflows, where they execute, the data they process, and where they are stored, combining and extending a data lake provenance representation with ML-specific data representation, following standards and reusing existing representations, such as W3C PROV, PROV-Wf, PROVLake, and MLS.

\noindent
\createDDP{dp:prospective-retrospective}{Keeping Prospection and Retrospection Related but Separated}
Davidson and Freire explain that prospective provenance captures the specification of a workflow, \ie{} the recipe of which data transformations will be processed and their inputs and outputs. In contrast, retrospective provenance captures the data that was consumed and produced, along with a detailed execution log about the computational tasks and execution environment~\cite{freire2008}.
The prospective provenance provides the abstraction layer to specify provenance analyses, often giving semantics to the retrospective provenance data generated during the workflows' execution. Also, there are cases that the provenance analysis uses only one kind of provenance data. 
Therefore managing both kinds of provenance data, and more importantly, with a strong connection between each related kind, is essential for the \MLView{}, which should be reflected in the provenance data modeling.

\begin{figure}[ht]
  \centering
  \includegraphics[width=0.7\columnwidth,trim={0 3.0cm 0 0},clip]{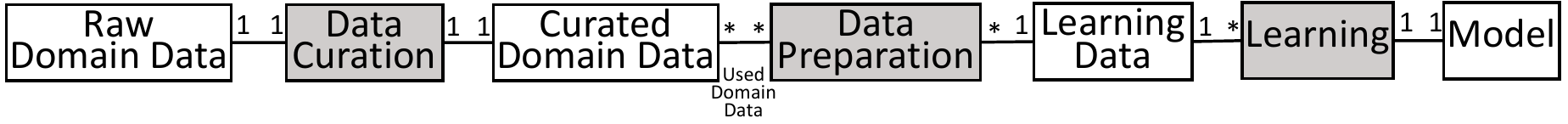}
  \caption{Conceptual data schema of the key concepts of the lifecycle and their supporting graphs.}
  \label{fig:simplemodel}
  \ourvspace{}
\end{figure}

\noindent
\createDDP{dp:conceptual-model}{Designing a Focused Conceptual Data Schema} 
To provide the specialized semantics needed by the \MLView{}, we propose a conceptual data schema focusing on the key concepts identified by the characterization in Section \ref{sec:ml_lifecycle}. The concepts are driven by the lifecycle phases and the data they manipulate:
the phases are illustrated with a gray background and the and four main kinds of data are illustrated in white background in the UML class diagram in Figure \ref{fig:simplemodel}.

On the four data concept classes, each instance represents one dataset, \ie{} a set of data elements that combined form one meaningful set of data for a given application.
As with any dataset, we may have a data schema that varies depending on the application; it may be further decomposed into several interrelated sub-datasets (or subconcepts for a given application), and there may be related metadata such as where it is physically stored and data sizes.
Concerning the three phases' classes, each can be further decomposed into workflows with associated execution data.
A \codefont{Learning} instance can be qualified into training, validation, and evaluation.
With respect to relationships, each \codefont{Data Curation} instance consumes a \codefont{Raw Domain Data} instance and generates a \codefont{Curated Domain Data} instance. Then, each \codefont{Curated Domain Data} instance may be consumed by one or more \codefont{Data Preparation} instances, which in turn may generate one or multiple \codefont{Curated Domain Data} instances (\ie{} a n:m relationship). For instance, a learning algorithm may require the preparation of well log data and seismic data, jointly, and thus two sets of \codefont{Curated Domain Data} would need to be related to the \codefont{Data Preparation} instance.
Finally, each \codefont{Data Preparation} instance generates a \codefont{Learning Data} instance to be be consumed by one \codefont{Learning} process that generates one \codefont{Model} instance.
Typically, during a learning phase, there are multiple \codefont{Learning} instances, each generating a \codefont{Model} instance.

\ourvspace{}
\subsection{\revisedcolor{System Architecture Design}}
\label{sec:system-design-principles}

\noindent
\createSDP{dp:portable-capture}{Portable and Distributed Capture Control}
As discussed, the workflows execute in highly distributed, heterogeneous environments processing data in heterogeneous data stores, executing within heterogeneous software and on heterogeneous environments.
To address this distributed execution control, the provenance system should be portable with distributed capture control so that there may be multiple provenance data capturers spread out across the multiple workflows executing. 
To address the heterogeneity of how workflows are executed, 
the provenance system cannot be tightly coupled with a specific workflow tool, but rather it should be
pluggable to any of these aforementioned heterogeneous ways of executing workflows. The distributed captured data are ultimately integrated into the unified \MLView{}.

\noindent
\createSDP{dp:microservices}{Specialized Microservices in a Distributed Architecture}
In addition to the distributed capture control, designing a provenance system using a microservices architecture
allows for the flexibility needed for large-scale deployments in hybrid environments. The provenance system can be decomposed into smaller, stateless microservices with specialized functions and, more importantly, it enables that components of the provenance system architecture are deployed wherever best fits for the workflow having provenance being captured.
For instance, provenance capture components can be deployed geographically near (or inside) the machine where the workflow runs, to reduce latency caused by communication costs, and other heavy-weight provenance-specific processes (\eg{} creating the linkages, inserting in the DBMS) and the DBMS itself can be deployed elsewhere, to reduce concurrency with the running workflows. A real deployment exploring the flexibility to place the architectural components to reduce communication costs and concurrency is shown in Section \ref{sec:implementation}.

\noindent
\createSDP{dp:scalable-capture}{Strategies for a Scalable Capture}
Since many of these workflows require HPC,
the provenance capture system should not add
significant performance penalties to the running workflows, requiring designing strategies for a scalable data capture. 
In addition to reducing concurrency, as described in \refSDP{dp:microservices}, which is one of these strategies, other strategies to reduce performance overhead are as follows.
\revisedcolor{During capture, the provenance persistence requests coming from the running workflows to the provenance system should be asynchronous and do not need to wait for complete processing, avoiding adding periods of waiting in the running workflow.}
Also, batches of data capture requests from the running workflows can be queued in the local memory of the host processing the workflow and sent to the provenance system at once, avoiding keeping open multiple communication channels between the running workflow and the provenance system. These batches are then received in the provenance system, which should process each request in the batch in a parallel manner, to reduce the time between the provenance capture in the workflow and the data to be readily available for queries in the \MLView{}.
Moreover, during capture, the provenance system responsible for creating the data linkages should avoid doing read operations to the underlying DBMS, but should only do appends to the data in the DBMS. This is because the read operations on the DBMS inevitably have to be waited for the query response, thus potentially increasing latency in the provenance capture. 
Finally, the only component that is in direct contact with a running workflow should be a lightweight provenance capture library shielding the workflows from possible slowness from other components. 
The key for such a lightweight library is to significantly reduce provenance-specific code in a workflow, consequently reducing provenance-specific calls during execution, and strictly follow the insert-only policy, so that no queries to the DBMS are made by the library, avoiding waits.
Such provenance-specific descriptions, essential for the specification of the workflows, are stored as prospective provenance data externally to the actual workflow. The provenance library (in the client-side of the system) does not need these specifications, which are essential for the server-side of the system, so the linkages that form the \MLView{} can be provided.
A side-effect of reducing provenance-calls in a workflow is that it also reduces the changes needed to be done, making it look as similar as possible to the original workflows without the hooks  \cite{provlake_escience_2019, thiago2020managing}.

\noindent
\createSDP{dp:unique-identifiers}{Easing Data Linkage with Unique Data Identifiers}
The concept of using unique identifiers is useful for keeping track of data in provenance systems~\cite{komadu_escience2016, koop2010bridging}.
Existing approaches keep track of data files consumed and produced in the workflows, and here we extend this concept to keep track of every data value that participates in the \MLView{}, even scalar values. Thus, every attribute-value pair that are consumed or produced in any data transformation participating in any workflow receives a unique identifier. 
So, whenever an attribute-value generated by one data transformation is consumed by another, the provenance system can reuse the value keeping track of the paths between transformations and, thus, keeping the workflows interconnected.

\noindent
\createSDP{dp:wf-design-hooks}{Workflow Design and Adding the Provenance Capture Hooks}
To enable the context-awareness (\refDDP{dp:context-awareness}), the first step
is to design the workflows with context awareness. For this, for each workflow in those multiple workflows for a given project, one needs to specify its data transformations with input datasets, parameters, and expected outputs.
Each computational process (data transformations) and the datasets they transform are qualified according to the \MLView{}'s conceptual data schema (\refDDP{dp:conceptual-model}).
When specifying data references, the physical location where the data reference is expected to be stored should be provided, as well as metadata about the execution environment where the workflow will execute. Finally, the relationships between the workflows and the data in the distributed data stores need to be specified.
Such specification can be maintained in configuration files, which will inform the provenance capture system to enable it to create the linkages to provide the context-aware integration of domain, ML, and hybrid environments and stores using provenance.
After the specification, hooks can be added to the workflows before and after each data transformation, informing the key concept (following the \MLView{}'s conceptual data schema) in each data transformation and data reference.
A data transformation execution is encapsulated by a provenance capture task, which typically occurs in a function call, a program execution, a web service call, or an iteration in an iterative workflow. 

\begin{figure}[ht]
  \centering
  \includegraphics[width=0.9\columnwidth]{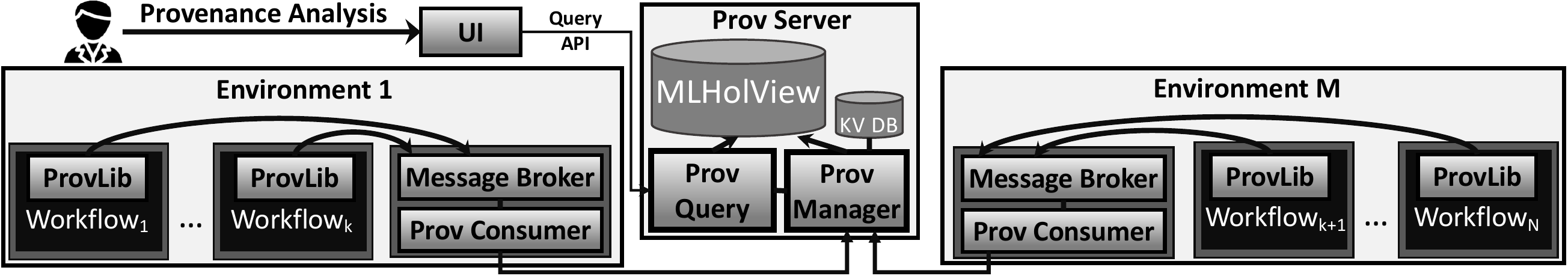}
  \caption{Reference system architecture to manage workflow provenance in the \MLCycle{}.}
   \label{fig:conceptual-arch}
\end{figure}

\noindent
\textbf{Reference System Architecture.} 
Based on these system design decisions, our proposed reference architecture is illustrated in Figure \ref{fig:conceptual-arch} and is described as follows.
There are $M$ environments (\eg{} HPC clusters, Kubernetes clusters) and $N$ workflows in all phases of the lifecycle, distributed on these environments.
Each workflow may use heterogeneous data stores and may be implemented as a standalone script, or as a workflow in a WMS, or a composition of microservice calls, or as a pipeline in a data processing or ML framework.
Provenance capture hooks, through a lightweight \codefont{ProvLib}, are added to capture provenance data at each data transformation in each of these workflows.
At the beginning and end of each (potentially parallel) data transformation executions for each (potentially parallel) workflow, a provenance capture event is emitted from the \codefont{ProvLib}. Thus a provenance capture event has the granularity of a data transformation execution, with their corresponding input data (at the beginning) and output data (at the end).
These events are asynchronously sent to a \codefont{Message Broker}, such as Apache Kafka, or any lightweight repository that persists these events in a queue.
Then, the \codefont{ProvConsumer}, which is a lightweight service that runs on the background, consumes from this 
queue and sends the requests to the \codefont{ProvManager}, which is aware of the prospective provenance data and can create the context-aware linkages using W3C PROV-based relationships and the reuse of unique identifiers (\refDDP{dp:unique-identifiers}), and sends the data to the \MLView{}, which is managed by a DBMS, typically a knowledge graph DBMS.
The (\codefont{Message Broker}, \codefont{ProvConsumer}) pair is instantiated at each environment to reduce communication costs between the ProvLib.
The ProvManager is a RESTful, stateless service and can receive provenance capture requests in any order. Thus it uses a lightweight Key Value DBMS (\eg{} Redis) to manage state when needed (\eg{} to create a link with a just received request with another request sent before). 
During the execution of these workflows, users or applications may submit provenance analysis through a Query API that communicates with the \codefont{Prov query} component, which is a RESTful service responsible for implementing query building strategies using the query language of the \MLView{}'s DBMS and returning the results to the requesting client.

\revisedcolor{
\noindent
\textbf{Analyzing Design Decisions Based on Existing Approaches.}
To the best of our knowledge, the design decisions 
\refDDP{dp:context-awareness} (integrating domain, ML, and hybrid environments and stores in a knowledge graph) and \refDDP{dp:provschema} (combining multi workflow, data lake, and ML provenance schemas) are new for general provenance management approaches. 
Regarding the decision \refDDP{dp:mlview} (holistic data view),
existing approaches \cite{souza2017data, silva_raw_2017} also propose provenance as a view over datasets, but cannot cope with data being generated by multiple workflows, such as the ones in the ML lifecycle; also, there is no ML-specific schema in their views.
The decision \refDDP{dp:prospective-retrospective} (prospective and retrospective provenance separation) is traditionally followed by most general provenance approaches, but it is not used in the existing approaches for ML provenance~\cite{esteves2015, publio2018, W3CML}. 
The system design decision \refSDP{dp:portable-capture} (portable and distributed control) is not often adopted by existing approaches \cite{miao_towards_2017,miao_sigmod2019}, as typically they propose ``all-in-one'' ML data management systems that require the workflows to be executed within such a platform and thus cannot address highly heterogeneous environments, parallel software, and data stores. The approaches that follow a more portable system design decision either cannot deal with multiple workflows \cite{silva_raw_2017} or does not follow the decisions \refDDP{dp:context-awareness} and \refDDP{dp:provschema} \cite{komadu_escience2016}. The other system design decisions \refSDP{dp:microservices}--\refSDP{dp:wf-design-hooks}
can be found in existing provenance-based approaches, nevertheless since they do not follow the data design decisions we are proposing, particularly \refDDP{dp:context-awareness} and \refDDP{dp:provschema}, they can only partially support the \MLCycle{}. 
In summary, we only found approaches that partially follow some of these decisions, which is not enough to answer complex end-to-end queries that integrate all lifecycle phases.
}

%% file: our-contents/sections/prov_data_representation.tex
\ourvspace{}

\section{
\revisedcolor{PROV-ML: A Provenance Data Representation For the \MLCycleUPPER{}}}
\label{sec:provml}

\begin{figure}[htb]
   \centering
\includegraphics[angle=-90,clip,trim={0.0cm 19cm 0.0cm 1cm}, width=0.51\linewidth,keepaspectratio]{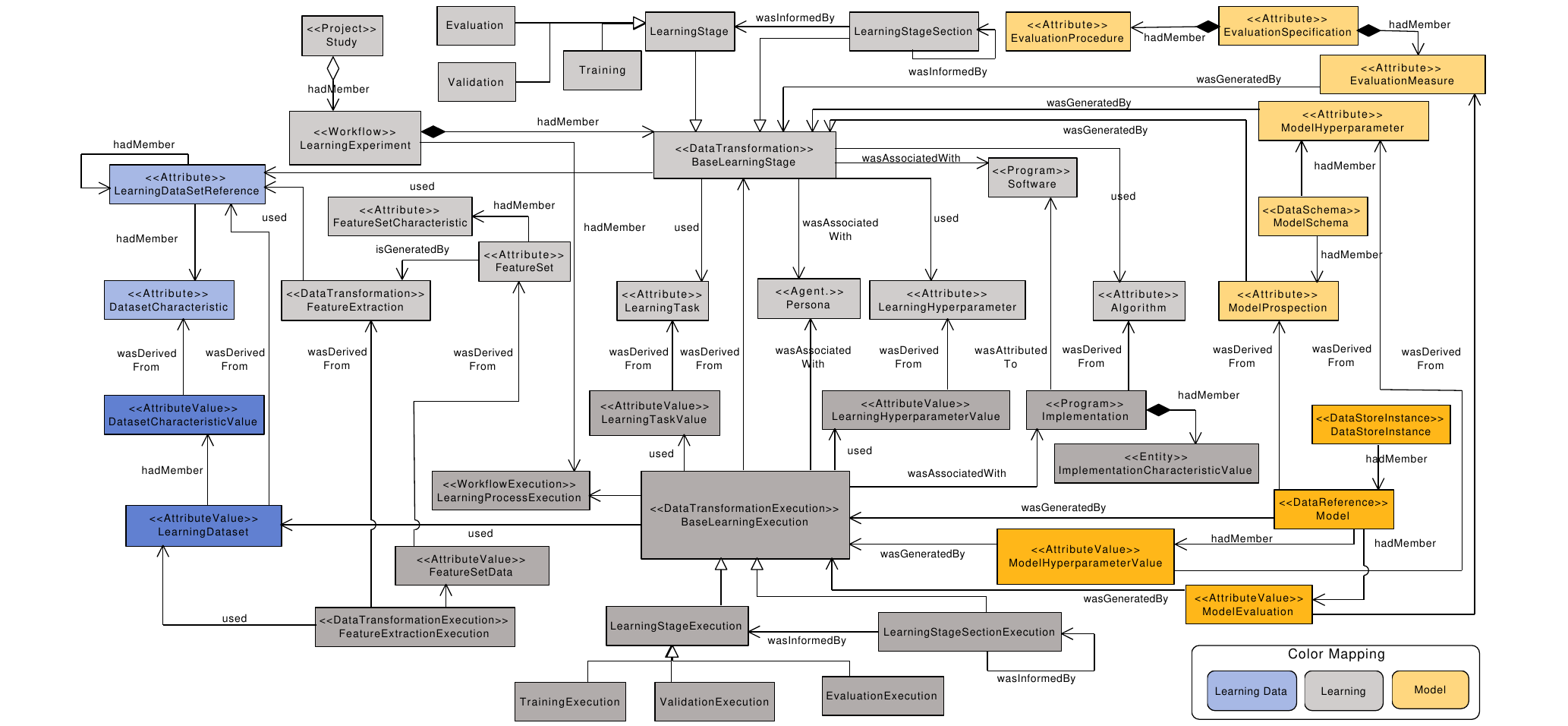}
   \caption{PROV-ML: a W3C PROV- and W3C ML Schema-compliant provenance data representation for scientific ML.}
   \label{fig:prov-lake-ml-model-data}
\end{figure}

In this section, we propose PROV-ML, the first generic provenance data representation for the \MLCycle{} to the best of our knowledge.
PROV-ML extends PROVLake \cite{provlake_web,provlake_escience_2019} as its underlying metamodel and employs elements of W3C ML Schema (MLS)~\cite{W3CML}. 
PROVLake is an extension of the W3C PROV data representation, specializing PROV for multiple workflows that process data in data lakes (see \refDDP{dp:provschema}).
However, as with other workflow provenance data representations,
PROVLake alone lacks ML-specific semantics. 
PROV-ML bridges this gap between workflows and ML concepts by building on PROVLake and MLS representations.
%
%
PROV-ML is depicted in Figure ~\ref{fig:prov-lake-ml-model-data}, where the light-color classes represent prospective provenance, and dark-color, retrospective. 
PROV-ML provides rich semantics and details based on the conceptual data model of the lifecycle's fundamental concepts (\refDDP{dp:conceptual-model}), especially the ones in the learning phase.
The colors in the figure map to these concepts: the blue-shaded classes account for the \codefont{Learning Data}; the gray-shaded, for the \codefont{Learning}; and the yellow-shaded, for the \codefont{Model}.
\revisedcolor{
The stereotypes indicated in the figure represent the classes inherited from PROVLake, which has subclasses that extend W3C PROV classes. 
This is how W3C PROV recommends how it should be extended\footnote{\url{https://www.w3.org/TR/prov-dm/\#section-prov-extended-mechanisms}}. 
All classes illustrated in the figure are individually described in Table~\ref{tab:provlakeml-classes}. We briefly discuss the PROV-ML classes here, and further details on the classes and on how PROVLake classes are extended from PROV classes are available online \cite{provlake_web}.
}

In PROV-ML, the \codefont{Study} class introduces a series of experiments, portrayed by the \codefont{LearningExperiment} class, which defines one of the three major phases in the lifecycle, the Learning phase. A learning experiment comprises a set of learning stages, represented by the \codefont{BaseLearningStage} class, which are the primary data transformation within the Learning phase and with whom the agent (\codefont{Persona} class) is associated.
The \codefont{BaseLearningStage} is as an abstract class from which \codefont{LearningStage} and \codefont{LearningStageSection} classes inherit. 
Also, it relates the ML algorithm, evoked through \codefont{Algorithm} class, used in the stage might be defined in the context of a specific ML task (\eg{} classification, regression), represented in the \codefont{LearningTask} class. 
This approach favors both the learning stage and learning stage section to conserve the relationships among other classes while grant them to have special characteristics discussed in the following. 
A learning stage varies regarding its type, \ie{} \codefont{Training}, \codefont{Validation}, and \codefont{Evaluation} classes. 
The provision of a specific class for the learning stage allows the explicit representation of the relationship between the Learning Data Preparation phase, through its Learning Data, and the Learning phase of an ML lifecycle. 
The \codefont{LearningStageSection} class introduces the sectioning semantics that grant capabilities of referencing subparts of the learning stage and the data, respectively. 
An example of sectioning elements relevance is the ability to reference a specific \textit{epoch} within a training stage, or mentioning a set of \textit{batches} within a specific epoch. The Learning Data appears in the model over the \codefont{LearningDataSetReference} class. Another data transformation specified in PROV-ML is the \codefont{Feature Extraction} class, which represents the process that transforms the learning dataset into a set of features, represented by \codefont{FeatureSet} class. This modeling favors the ML experiment to be reproducible since it relates the dataset with the feature extraction process and the resulting feature set. 

\input{our-contents/tables/provml_table.tex}

Further fundamental aspects regarding the Learning phase are the outputs and the parametrization used to produce these outputs. Like so, The \codefont{ModelSchema} class describes the characteristic of the models produced in a learning stage or learning stage section, such as the number of layers of a neural network or the number of trees in a random forest. The \codefont{ModelProspection} class represents the prospected ML models, \ie{} the reference for the ML models learned during a learning stage or learning stage section of a training stage. In addition to the data produced in the Learning phase is the \codefont{EvaluationMeasure} class. This class, combined with \codefont{EvaluationProcedure} and \codefont{EvaluationSpecification} classes, provide the representation of evaluation mechanisms of the produced ML models during any stage of learning, specifically: 
an evaluation measure defines an overall metric used to evaluate a learning stage (\eg{} accuracy, F1-score, area under the curve); 
an evaluation specification defines the set of evaluation measures used in the evaluation of learned models; and, 
an evaluation procedure serves as the model evaluation framework, \ie{} it details the evaluation process and used methods. On the parametrization aspect, PROV-ML afford two classes \codefont{LearningHyperparameter} and \codefont{ModelHyperparameter}. The first hyperparameter-related class represents the hyperparameter used in a learning stage or learning stage section (\eg{} max training epochs, weights initialization). The second class is used in the representation of the models' hyperparameters (\eg{} network weights).
Finally, PROV-ML addresses the retrospective counterpart of the classes mentioned above. The classes ending in \codefont{Execution} and \codefont{Value} are the derivative retrospective analogous of data transformations and the attributes, respectively. 

\revisedcolor{
Comparing with the paper\cite{souza_provml_2019} being extended here,
we make the following improvements in PROV-ML. We change the representation of learning stage types (\ie{} training, evaluation, and validation) from an enumeration to a hierarchy to make explicit relationships specific to learning phases. Also, towards better defining the stages within a learning experiment, we introduce the Learning Stage Section (\codefont{LearningStageSection} and \codefont{LearningStageSectionExecution} which is crucial to represent the specific characteristics sub-parts of a learning process, when it is required. Combining these representations, we enable the representation of specific events as sub-sections in a learning stage. For instance, in the training stage we now may represent mini-batch iterations, which is a common sub-section to enhance stochastic optimization~\cite{li2014minibatch}, and thereafter make references to the mini-batch representations. Likewise, we introduce the \codefont{Persona} class, subclass of PROV Agent, and associate it with \codefont{BaseLearningStage} and \codefont{BaseLearningExecution}. Further, we refined the names of elements related to hyperparameters and model besides description adjustments due to learning stage section creation.}





%% file: our-contents/tables/provml_table.tex
\begin{table}[ht]
\footnotesize
\caption{PROV-ML data representation classes.}
\topsepremove
 \begin{tabular}{>{\arraybackslash}m{0.20\textwidth}m{0.80\textwidth}}
\rowcolor{table_header}
\textbf{Class} & \textbf{Description} \\
\toprule
Study & Investigation (\eg{} research hypothesis) leading ML workflow definitions.
\\ 
\rowcolor{table_row_even}
LearningExperiment & The set of analyses (\eg{} research questions) that drives the ML workflow. 
\\ 
\rowcolor{table_row_odd}
LearningProcessExecution & An ML workflow execution. This is equivalent to \codefont{mls:Run} and was renamed to explicitly preserve the aspects of retrospective provenance, which are not explicitly handled in MLS. 
\\
\rowcolor{table_row_even}
LearningTask and LearningTaskValue & Defines the goal of a learning process, \ie{} the ML task  (\eg{} \codefont{LearningTask}: \textit{Classification}; \codefont{LearningTaskValue}: \textit{Seismic Stratigraphic Classification}). 
\\ 
\rowcolor{table_row_odd}
BaseLearningStage and BaseLearningStageExecution & Abstract classes of \codefont{LearningStage} and \codefont{LearningStageSection}, and their execution counterparts. It is used to conserve the relationships among other classes while granting them to have special characteristics.
\\
\rowcolor{table_row_even}
Persona & The personas associated with the learning process.
\\
\rowcolor{table_row_odd}
LearningStage and LearningStageExecution & Defines a stage in the learning process (\codefont{Training} or \codefont{Validation} or \codefont{Evaluation}) and its execution.
\\ 
\rowcolor{table_row_even}
LearningStageSection \  and \newline 
LearningSectionExecution & Introduces the sectioning semantics, \ie{} capabilities for provenance of subparts of the learning stage and corresponding data.
\\ 
\rowcolor{table_row_odd}
LearningDatasetReference and LearningDataset & Defines the dataset to be used by a \codefont{LearningStage} or \codefont{LearningStageSection}. In the last case, it is a section of a \codefont{LearningDatasetReference}. \codefont{LearningDataset} is the dataset used in the execution.
\\ 
\rowcolor{table_row_even}
DatasetCharacteristic and DatasetCharacteristcValue & Defines metadata about the \codefont{LearningDatasetReference} (\eg{} \#instances),  and \codefont{DatasetCharacteristcValue} relates with a \codefont{LearningDataset} (\eg{} \#instances =8).                      
\\ 
\rowcolor{table_row_odd}
FeatureExtraction and FeatureExtractionExecution & Defines the prospective and retrospective feature retrieval process, respectively.  
\\
\rowcolor{table_row_even}
FeatureSet and FeatureSetData & Defines the features and feature values a \codefont{FeatureExtraction} should generate over  \codefont{LearningDatasetReference}.
\\ 
\rowcolor{table_row_odd}
FeatureSetCharacteristic  & Defines the set of metadata that describes the \codefont{FeatureSet} (\eg{} number of features, features' type).                                  
\\ 
\rowcolor{table_row_even}
Software & Defines a collection of ML techniques' implementations (\eg{} Scikit-Learn).                                                 
\\ 
\rowcolor{table_row_odd}
Algorithm & ML technique with no associated technology, software or implementation (\eg{} k-means clustering technique).        
\\ 
\rowcolor{table_row_even}
Implementation & Defines the retrospective aspect of an \codefont{Algorithm}, \ie{} an ML technique's implementation in a software (\eg{} Scikit-Learn's k-means implementation).                               
\\ 
\rowcolor{table_row_odd}
ImplementationCharacteristicValue & Defines the implementation's set of metadata (properties and values),\eg{} version, git hash.    
\\ 
\rowcolor{table_row_even}
LearningHyperparameter & Defines the prior parameter of an \codefont{Algorithm} used by a \codefont{LearningStage} or \codefont{LearningStageSection}.
\\ 
\rowcolor{table_row_odd}
LearningHyperparameterValue & Defines the parameter values of an execution (\eg{} the $k$ value in a k-means clustering technique, range of epochs in a neural network training).                                   
\\ 
\rowcolor{table_row_even}
ModelSchema & The scope of the resulting model.                                                         
\\ 
\rowcolor{table_row_odd}
ModelProspection and Model & The resulting model a \codefont{LearningStage} or a \codefont{LearningStageSection} should generate, and the generated value (\eg{} the trained model after the training stage).                                                                       
\\ 
\rowcolor{table_row_even}
ModelHyperparameter and ModelHyperparameterValue    &  Hyperparameters a \codefont{LearningStage} or a \codefont{LearningStageSection} generate, and their values corresponding to the resulting model (\eg{} the epoch which the resulting model was generated).               
\\ 
\rowcolor{table_row_odd}
DataStoreInstance & Storage of the resulting model.                                    
\\ 
\rowcolor{table_row_even}
EvaluationMeasure and ModelEvaluation & A measure a \codefont{LearningStage} or a \codefont{LearningStageSection} should evaluate and the generated value (\eg{} the precision of classifier model).                                                                                
\\ 
\rowcolor{table_row_odd}
EvaluationSpecification and EvaluationProcedure & Classes directly inherited from MLS, with their semantics preserved. \\
\arrayrulecolor{table_row_odd}\hline
\end{tabular}
\label{tab:provlakeml-classes}
\end{table}

%% file: our-contents/sections/experiments.tex
\ourvspace{}
\section{SYSTEM IMPLEMENTATION AND EXPERIMENTAL EVALUATION}
\label{sec:exps}

In this section, we provide experimental validation of the design decisions to build and query the \MLView{} to support the  \MLCycle{} in a real case study in the O\&G industry.
First, we explain how we implement and deploy the system used in the evaluation (Sec. \ref{sec:implementation}). Then, we show a running example of which data are captured during execution of the workflows to answer the exemplary queries \queries{} (Sec. \ref{subsec:usecase}).
After, we present performance and scalability analyses of the system 
(Sec. \ref{subsec:performance_analysis}). 
Then, we discuss the benefits of PROV-ML both in terms of easing queries and query performance (Sec. \ref{subsec:exp-query-comparisons}).
Finally, we discuss how our approach can be customized (Sec. \ref{subsec:customization}) and conclude with lessons learned after this evaluation (Sec.  \ref{sec:lessons-learned}). 

\ourvspace{}
\subsection{Implementation and Deployment}
\label{sec:implementation}

ProvLake \cite{provlake_web} is a provenance system capable of capturing, integrating, and querying data across distributed services, programs, scripts, and data stores used by multiple computational workflows using provenance data management techniques \cite{provlake_escience_2019, souza_provml_2019}. 
In this section, we explain how we implement the design decisions to enable ProvLake to build the \MLView{} and how it is deployed to support the lifecycle in our case study.

\smallskip
\noindent \textbf{ProvLake Architecture.} 
ProvLake architecture is an implementation of the reference architecture (Fig. \ref{fig:conceptual-arch}).
Details about this architecture can be found in our previous work \cite{souza_provml_2019}. Here we give a summary, highlighting how its components are mapped to the reference architecture proposed in this paper. 
The ProvLake Library (PLLib) \cite{provlake_github} maps to the ProvLib.
ProvTracker implements simple queue management to receive the provenance capture events coming from the lib and also implements a queue consumer, thus working both as the message broker and the provenance consumer in the reference architecture. ProvManager maps like the reference architecture and the
PolyProvQueryEngine is the component for building the provenance queries and sending them to the DBMS managing the \MLView{}.
As described in the decision~\refSDP{dp:wf-design-hooks},
the workflows are specified using prospective provenance data stored as configuration files.
Data transformations that are specific and standard in ML workflows, \eg{} training, validation, and evaluation are defined beforehand
following the conceptual data schema for the key concepts (\refDDP{dp:conceptual-model}) and the PROV-ML  (Sec. \ref{sec:provml}) for attributes, such as hyperparameters and model evaluation attributes.
ProvTracker uses the specified prospective provenance data to provide for the tracking by creating the relationships of retrospective provenance data being continuously sent by PLLib added to the workflows.
ProvTracker gives unique identifiers (\refSDP{dp:unique-identifiers}) to every data value captured and when there are data references (\eg{} references to files or identifiers in a database table or any analogous data reference), it creates a knowledge graph relationship between the data value and the data store \cite{provlake_escience_2019}. ProvManager transforms the captured data into RDF triples (the data model of the DBMS in use by ProvLake in this implementation) 
following the PROV-ML ontology (when capturing data in the learning phase) and PROVLake ontology (when capturing data in the previous phases of the lifecycle).

\smallskip
\noindent \textbf{ProvLake Deployment in the Case Study.} The deployment of our case study also follows the system design decisions (Sec. \ref{sec:system-design-principles}).
It uses two clusters: a Kubernetes cloud cluster for data curation and learning data preparation workflows, and the other is a large HPC cluster with CPUs and GPUs for the workflows in the learning phase. 
PLLib is the only component in direct contact with the users' workflows running in the clusters (\refSDP{dp:scalable-capture}).
This deployment is illustrated in detail in our previous paper \cite{souza_provml_2019}.


\smallskip
\noindent \textbf{Hardware Setup.} The experiments use three environments. An HPC cluster for learning workflows, which has 393 Intel and Power8 nodes, each with 24 to 48 CPU cores, 256 to 512 GB RAM, interconnected via InfiniBand, sharing about 3.45 PB in a GPFS, and using in total 946 GPUs (NVIDIA Tesla K40 and K80, each with 2880 and 4992 CUDA cores respectively); a Kubernetes cloud cluster for data processing, which has 4 nodes, each with 16 GB RAM and 8 cores; and a server machine Intel Core i7-7700T CPU 2.40 GHz, 8 GB DDR4 RAM, 128 GB SSD Liteon.

\smallskip
\noindent \textbf{Software Setup.} ProvManager, PolyProvQueryEngine, and Prov DBMS are deployed on a virtual Kubernetes cluster with two nodes with 4 vCores, 16 GB RAM each, virtualized on top of the data processing cluster. ProvManager's queue is set to 50, and ProvTracker threads are set to 120. The workflow scripts of our use case are implemented in Python using different libraries to manipulate raw seismic files and for designing and training the ML algorithms (PyTorch V1.1) that execute in the HPC cluster.
For the query performance tests, we deployed three different DBMSs on the server machine: Apache Jena TDB 3.12, Allegro 6.6.0, and Blazegraph 2.1.5.

\ourvspace{}
\subsection{Use Case Validation} 
\label{subsec:usecase}

\begin{figure}[h]
  \centering
  \includegraphics[width=0.9\textwidth,keepaspectratio]{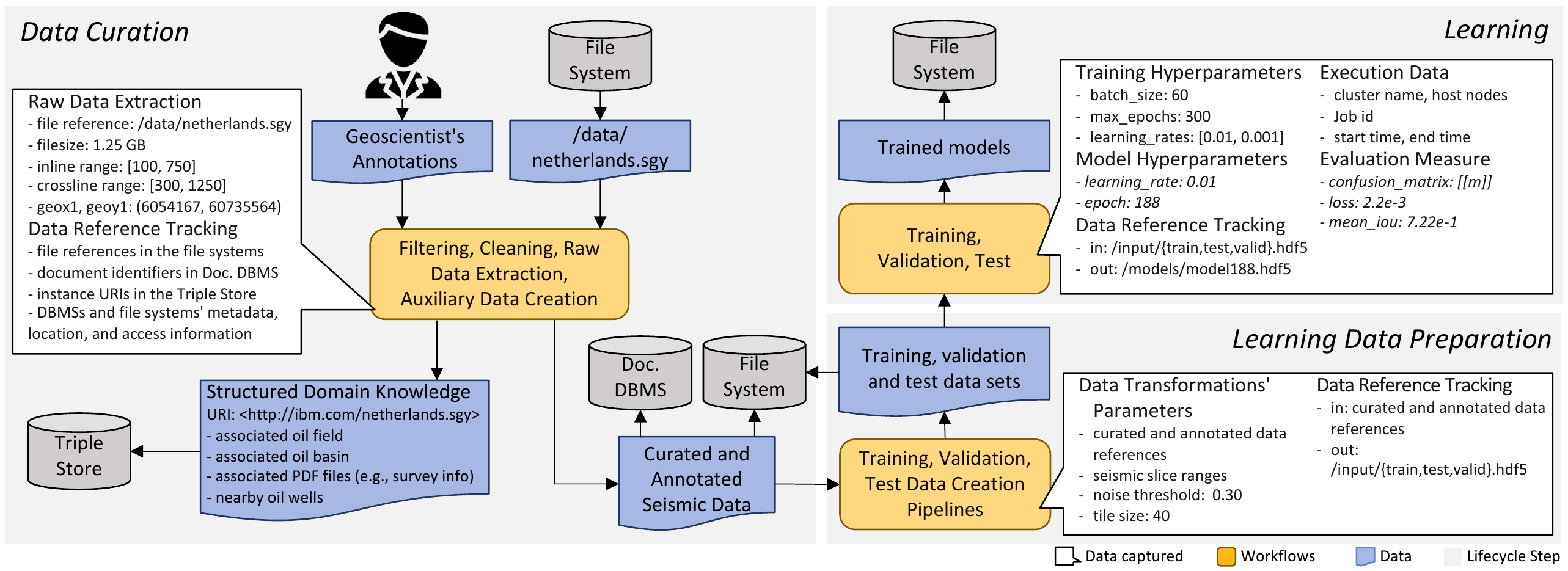}
  \caption{Summarized example of provenance tracking in an O\&G use case.
  Details on the captured data, contents, stores, and the dataflow used to answer the queries \queries{} are in Table~\ref{tab:use-case}.}
   \label{fig:usecase}
\end{figure}

In this section, we investigate whether our approach supports the lifecycle by enabling users to perform comprehensive, \ie{} end-to-end analyses that integrate the data consumed and generated in the workflows, from raw domain data to learned models.
More specifically, we investigate if the proposed data design decisions (Sec. \ref{subsec:data_designprinciples}) can be applied to answer queries that do such integration of the data.
We explore the O\&G use case described in Section~\ref{sec:ml_lifecycle} and validate if the data tracked by ProvLake, inserted in the \MLView{} implementing the PROV-ML data, can answer the queries \queries{}. 
Fig.~\ref{fig:usecase} shows the phases of the lifecycle in this use case. Next, we describe the workflows of the use case and how ProvLake tracks the data.


In the data curation phase, ProvLake tracks provenance while data-intensive scripts run.
When processing raw files, essential data that will help answer the queries are extracted, associated with the file's URI, and stored in the provenance database.
One example of such data is the embedded geographic coordinates in raw SEG-Y seismic files.
Additionally, geoscientists add relevant information, based on their specialized knowledge, as input to some of those scripts to be loaded into a domain-specific knowledge graph database, external to the provenance database, but also tracked by ProvLake through links between the workflows and the domain knowledge in the graph. Relevant information includes associated oil fields, basins, oil wells, and pieces of text from PDF documents with survey information related to the geological data acquisition process.
These annotations are stored in triple stores in a domain-specific database, externally to the provenance database.

The learning data preparation phase includes several data transformations in a pipeline that converts the curated and annotated scientific data into training, validation, and evaluation datasets.
Each transformation contains parameters that specify, for instance, noise filter thresholds, input shape, or the selected seismic lines (inlines, or crosslines) of the seismic cube that constitutes the training dataset.
Each value of these parameters, the name of the transformation, the execution data, and the references to input and output data are captured and represented in ProvLake's provenance data graph.

The entire process is interconnected, where each phase produces data and passes it forward for the consumption of the next one.
Essentially, ProvLake tracks and maintains such interconnections in a provenance data graph composed of RDF triples.
Such structures describe chained data transformations in the multiple workflows that constitute the inner phases of the major ones of the lifecycle run.
RDF resources represent the data in Fig.~\ref{fig:usecase}, \ie{} instances that extend \codefont{prov:Entity} and PROV-ML specializations.
Each of these instances receives a URI, which works as a global identifier throughout the lifecycle \refDDP{dp:unique-identifiers}.
Examples of RDF resources are learned models produced in the learning phase, a model's hyperparameters, evaluation metrics, and references (file path) to actual model files stored in the file system.
Provenance data graphs also associate execution data with learned models.
Execution data may include file system metadata, a cluster's hostname and node names used in the HPC jobs, job ids in the cluster scheduler, or start and end timestamps of each block of provenance capture events.

ProvLake can keep track of data distributed in multiple stores. Such ability helps to maintain data relationships between raw files in the file system and structured knowledge stored in another database.
Auxiliary data, such as polygons in the seismic cube, are stored in the Document DBMS. The system similarly tracks data references and related to the raw files.
Other data, such as implementation details, software name, and version, are captured and stored in the provenance database, following the PROV-ML, but, for simplicity, we do not show them in the figure.
Finally, since the system tracks every data and their relationships while the workflows execute, ProvLake enables answering online, offline, intra- and inter-training provenance queries to analyze ML data, domain-specific data, and execution data throughout the phases of the lifecycle, exemplified by the queries \queries{}.

To submit queries, the user sends a GET or POST request to one of PolyProvQueryEngine's endpoints.
Then, PolyProvQueryEngine sends requests to ProvManager.
Most of the queries are answered with simple graph traversals using standard SPARQL features.
For instance, to answer Q1, the user provides a learned model URI (generated in the learning phase), and the query should traverse in the provenance data graph backward until the raw seismic file's URI (processed in the data curation phase).
One can get the geographic coordinates and number of seismic slices by querying the extracted data related to the seismic file.
In turn, to obtain the oil basin and oil field information, the query retrieves data from the resource, in the Triple Store, which represents structured knowledge about the seismic file.
For Q2 and Q6, one can execute a similar graph traversal.
Other queries require analytical operators, such as Q3, which requires finding the learned model with least (using \codefont{min()} native SPARQL operator) loss and returning its hyperparameters.
Q4 and Q5 make use of execution data to provide basic statistics (\codefont{min(), max(), avg()} operators) about the execution time of training iterations and Q7 retrieves the models, their hyperparameters, their evaluation measures, and minimum batch loss per model generated when a specific learning dataset was used.

\ourvspace{}
\subsection{Performance Analysis on Capturing Data during Learning Workflows}
\label{subsec:performance_analysis}

In our use case for training an autonomous identifier of geological structures (\emph{c.f.} Sec \ref{sec:ml_lifecycle}), the learning phase generates a large amount of provenance data at a high frequency to stress ProvLake services. In the deep learning model training, there are two provenance capture calls (for the beginning and end) at each batch iteration in each learning epoch. In this test, each learning workflow executes about 35 iterations for each learning epoch and up to 300 epochs, generating about 15,000 provenance capture events per workflow run. ProvTracker runs on one node in the learning cluster with 24 CPU cores, whereas the learning workflows run in parallel and distributed on up to 8 nodes, each with 28 Intel CPU cores and 6 GPUs (K80). While running the workflows, PLLib captures data at runtime and sends them to ProvTracker, which in turn sends them to ProvManager service deployed externally on the virtual Kubernetes cluster, which finally stores them in the Prov DBMS. A provenance capture overhead analysis of ProvLake using synthetic workloads to highly stress the system and comparison with a competing system has been presented in a previous work \cite{provlake_escience_2019}. Here, we  evaluate the system design decisions that focus on providing distributed capture control and scalable architecture (\refSDP{dp:portable-capture}--\refSDP{dp:scalable-capture}). We test different settings for provenance analysis and then test the scalability using real ML workloads in both cases.
We measure the overall execution time of the learning workflow script, repeating each test at least 10 times and we plot the boxplots of the repetitions and the numeric values used in-text refer to the medians.

\noindent \textbf{Varying Provenance Capture Settings.}
The PLLib allows customizing provenance capture settings, such as the queue size and whether the provenance capture events should be persisted to the local disk, rather than sending to ProvTracker. Then, if disk only is not specified, when the scripts execute, provenance data are captured and sent to ProvTracker.

\begin{figure}
\centering
\begin{subfigure}{.35\textwidth}
\captionsetup{size=normalsize}
  \centering
    \includegraphics[width=\textwidth]{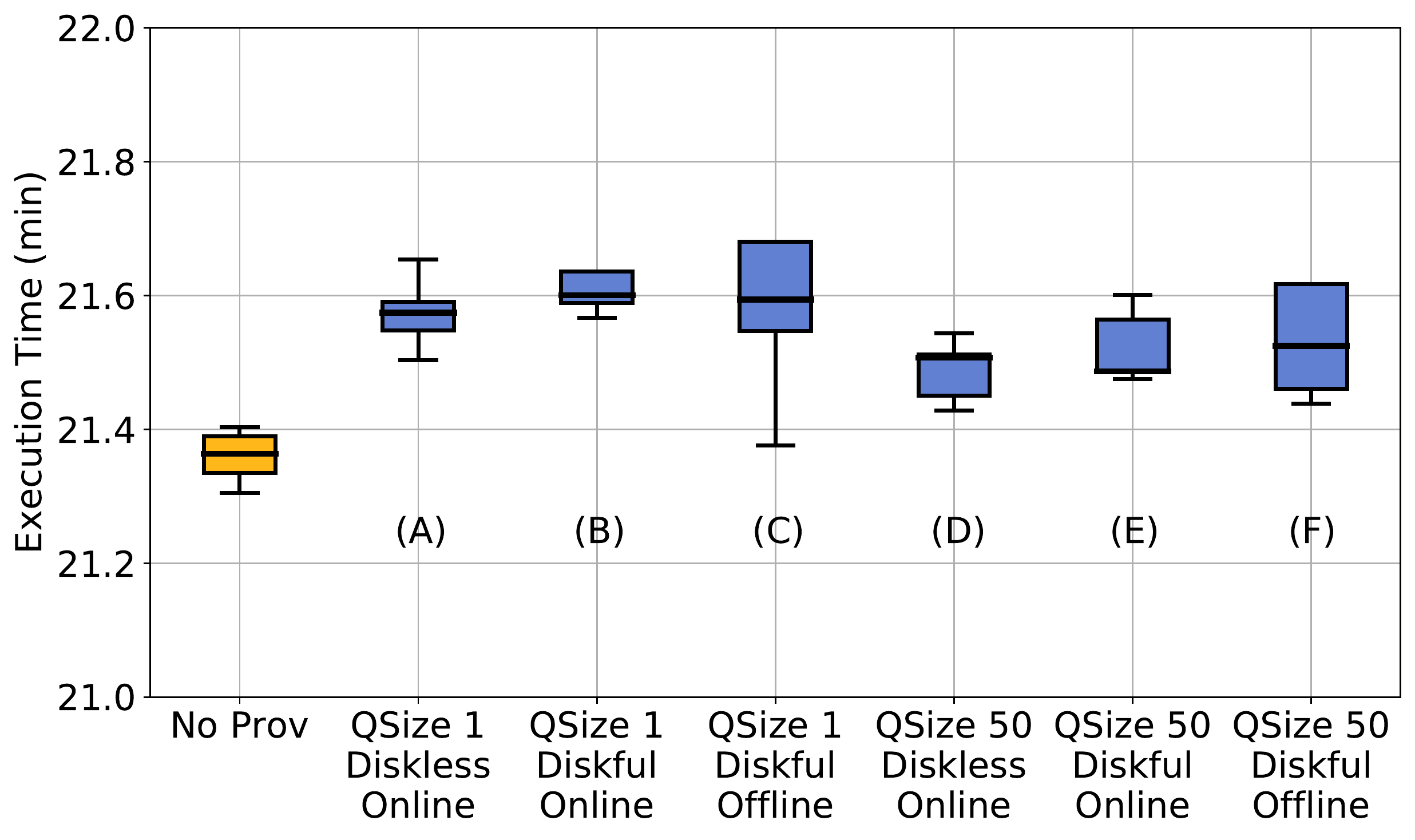}
   \caption{}
   \vspace{-1em}
\end{subfigure}%
\begin{subfigure}{.10\textwidth}
\end{subfigure}
\begin{subfigure}{.35\textwidth}
\captionsetup{size=normalsize}
  \centering
   \includegraphics[width=\textwidth]{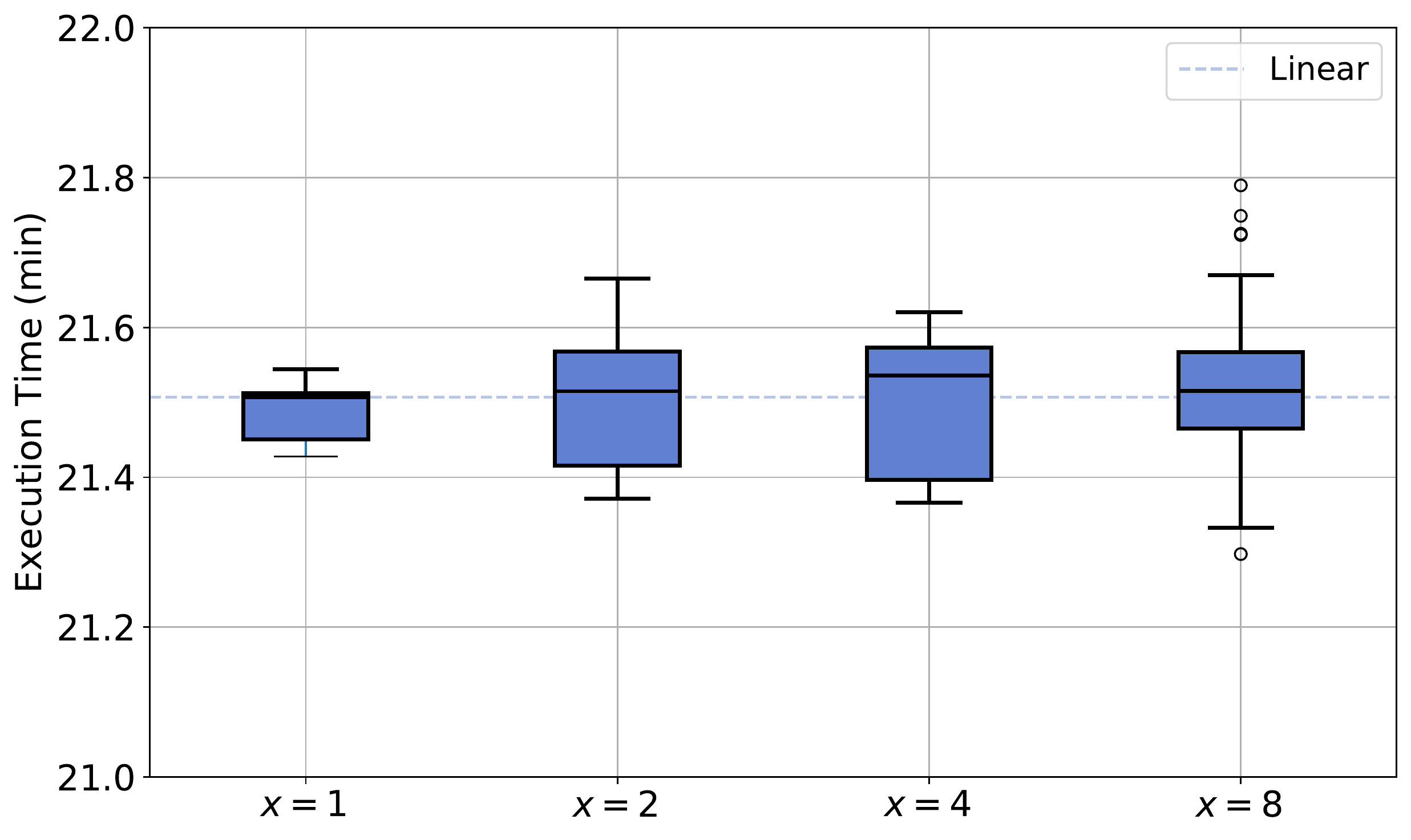}
   \caption{}
   \vspace{-1em}
\end{subfigure}

  \caption{Performance analysis results.
  Figure \ref{fig:perf_analysis}~(a) shows the variation of provencance capture settings, where Setting D adds 0.67\% overhead.  
  Figure \ref{fig:perf_analysis}~(b) shows the scalability results, a near-linear scalability with up to 48 GPUs and 228 CPUs.}
  \label{fig:perf_analysis}
  \ourvspace{}
\end{figure}

For a baseline, we first execute the training without any provenance capture, then we vary the queue size in PLLib (\ie{} amount of provenance capture requests accumulated in PLLib), diskless vs. diskful (\ie{} saving or not provenance data in a log file on disk), and online vs. offline (\ie{} storing or not provenance data in the DBMS, available for online provenance queries during the execution). As for the training datasets, we use a curated and labeled real seismic dataset using a specific range of seismic slices (corresponding to a regional section of a seismic cube) defined by the model trainer. The results are in Fig.~\ref{fig:perf_analysis}~(a), where the fastest result is for Queue Size = 50, Diskless, Online (Setting D). Comparing with the setting with no provenance capture, the added execution overhead, in this case, is only 8.6 seconds on top of 21.3 minutes, \ie{} 0.67\%, which is considered negligible. 
\revisedcolor{Although these workflows execute in parallel in a total of around 21.3 minutes, in this experiment, there are 8 workflows concurrently running using 228 CPUs and 48 GPUs. 
It is only a snapshot of the whole lifecycle, whereas in practice, this set of 8 (or more) workflows needs to be executed hundreds of times as users adjust the experiment setting and assess the model performance. They only stop when they are satisfied with the results, which may take weeks of experimentation.
Also, this is a critical part of the lifecycle, responsible for stressing the provenance capture system while training in an HPC cluster. 
}

To analyze the queue size, we compare Settings A--C with D--F, and we see larger queues provide faster provenance capture since there is less but larger communication with ProvTracker service. For instance, Setting A is about 7\% slower than D.
\revisedcolor{Persistence latency (\ie{} time taken between data capture in the running workflow and the actual persistence in the database) is not the focus of this experiment, as we are interested in understanding whether our approach adds significant capture overhead, which is what impacts the running user workflow performance. However, we briefly discuss it here as very large queues can introduce higher persistence latency and may impact the user experience, especially for monitoring. This latency may occur for many reasons, including network traffic, from the running workflow to the DBMS passing through the provenance system, and queue throttling at the DBMS receiving the persistence events, as DBMSs typically also implement queuing mechanisms. 
In the settings with queue size 50 (D--F), a persistence latency of less than 5 seconds was empirically observed, which is considered good enough for training monitoring.} To analyze diskless vs. diskful settings, we compare Setting A with B and C; and D with E and F. Diskless is faster than diskful, as the latter introduces more I/O operations at runtime. However, comparing only the medians, the difference is negligible (less than 0.1\%). Thus, because of a higher fault tolerance provided by a diskful setting, it may be useful to append provenance data onto a file on disk, locally in the cluster where the workflow runs. 
Similarly, comparing the medians, we observe that the difference between online vs. offline (\eg{} setting B vs. C or E vs. F) is also small, about 1\%. Therefore, despite (D) being the fastest setting, (E) may be preferred because its performance is nearly the same as (D), and it has the advantage of backup storage for provenance data, which is quite important as provenance is used for reproducibility. 

\smallskip
\noindent \textbf{Scalability Analysis.} In this experiment, we want to confirm if the execution strategies on an HPC cluster are keeping the overhead low in a real ML workload, running multiple learning workflows in parallel. We run a weak scalability test by increasing the number of processing units while increasing the data size. We use the fastest setting of the previous experiment (\ie{} D) and the same seismic cube. To set up the training datasets, the trainer selects up to 8 different sets of seismic slices, where each set has the same length (\ie{} nearly the same data size). Thus, for $x \in \{1, 2, 4, 8\}$, there are $x$ workflows running on $x$ nodes in parallel, summing $28x$ Intel CPU cores, $6x$ GPUs, $4992*6x$ CUDA GPU cores, using in total an input dataset with size $x*datasize$, where $datasize$ is the size of a dataset formed by 1 set of seismic slices. The results are in Fig.\ref{fig:perf_analysis}~(b), where we illustrate the linear scalability as a horizontal line passing through the median of the smallest setting ($x=1$). Ideally, the medians should be near this line. If they are not, it means that ProvTracker is taking too long to answer, caused by high stress in the system due to too many provenance capture requests, adding latency to the training. However, we see that even in the largest setting (\ie{}  $x=8$), the execution time remains close to the linear curve. The boxes remain within a small margin of 0.2 min (or 0.9\% of the $x=1$ median) between 21.4 and 21.6 min, meaning that the system delivers a constant and predictable behavior even at larger scales. We note though that the variance grows with the scale, caused by the larger number of parallel tasks. Therefore, we conclude that at least for this scale (up to 48 K80 GPUs), the provenance capture system delivers good scalability.

\ourvspace{}
\subsection{Query Comparisons With and Without PROV-ML}
\label{subsec:exp-query-comparisons}

In this experiment, we analyze the benefits of PROV-ML, both qualitatively and quantitatively.
We begin with a qualitative comparison of queries that use PROV-ML, highlighting its expressiveness and the complexity of building queries that use or not PROV-ML. Then, we further provide a quantitative analysis to investigate whether using PROV-ML can help to accelerate queries and, if it can, how much it helps. 
Among the queries \queries{}, we select three to compare in detail: Q1, Q5, Q7, and the reason for this choice is that they increase in complexity and how much they make use of the concepts modeled specifically in the PROV-ML ontology (\ie{} emphasis on the learning phase, \textit{c.f.} Sec. \ref{sec:provml}).
Q1 is the simplest query and makes the least use of PROV-ML specific concepts, Q7 is the most complex query with the heaviest use of PROV-ML, and Q5 is in between these two.
We write the selected queries both with and without the PROV-ML ontology (written in OWL) using SPARQL 1.1. The query complexity stems from the number of clauses to filter, patterns to match in the graph traversal, aggregations and sorting, and amount of triples that satisfy the patterns to match; and the number of clauses that make use of the PROV-ML ontology defines how much each query makes use of the PROV-ML ontology. 
\todo[]{tentar colocar essas queries num github publico}

\noindent
\textbf{Qualitative comparison.} 
Since Q7 is the most complex query and makes heavy use of PROV-ML, it helps us to illustrate whether PROV-ML eases query building, especially when there is heavy use of Learning phase concepts.
Excerpts of Q7 in SPARQL with and without PROV-ML are available in the Listings \ref{listing:q7-provml} and \ref{listing:q7-noprovml}, respectively.
\todo[]{no caption das listings, colocar uma mencao pro github com as queries completas}
Comparing both, since PROV-ML has specialized concepts for the Learning phase, it requires fewer clauses to be matched to express the same concept. 
For instance, to match triples in the training stage only, with PROV-ML, we just write one clause (Lst. \ref{listing:q7-provml}\#L2), whereas without PROV-ML we need to write four clauses to qualify the data transformation needed for the query (Lst. \ref{listing:q7-noprovml}\#L1--6).
This is because there are other stages (evaluation, validation), and since without PROV-ML there are no specific types for each stage, we need to qualify the variable \codefont{?training} to determine the correct stage.
Without PROV-ML, the only resource we have to do this is to tag the data transformations that are related to training. 
In PROVLake ontology, tagging of workflows, data transformations, and attributes is possible with the property \codefont{provlake:tag}, but since naming, schema definitions, and tagging are available only in the prospective part, we need three more  clauses: one to relate the retrospective instance with its prospective instance (Lst. \ref{listing:q7-noprovml}\#L4),
other to give the type of the prospective instance (\#L6), and a third clause to qualify the data transformation as ``Training'' using the tag property (\#L7).
A similar fact happens for the epoch iteration part of the query (Lst. \ref{listing:q7-provml}\#L3--6 and Lst. \ref{listing:q7-noprovml}\#L8--12) and also for the model reference (Lst. \ref{listing:q7-provml}\#L16--19 and Lst. \ref{listing:q7-noprovml}\#L26--33) and for model evaluation (Lst. \ref{listing:q7-provml}\#L20--24 and Lst. \ref{listing:q7-noprovml}\#L34--42).
However, the model hyperparameters part of Q7 differs from the others because Q7 requires the name of the hyperparameters, in addition to the values, and names are stored within the prospective portion of the data. In this case, Q7 demands a relationship between prospective and retrospective, regardless it is with PROV-ML or not. The only difference between with and without PROV-ML for such cases is that with PROV-ML, we do not need tags to qualify the attributes of interest (Lst. \ref{listing:q7-noprovml}\#L24, as we can specify hyperparameters using the specialized type (Lst. \ref{listing:q7-provml}\#L14). As a result, in these cases, only one extra clause is needed.

\todo{link gh}

Therefore, we found that when the parts of the query do not demand prospective provenance data, one needs to write three extra clauses when not using PROV-ML (a clause to relate the retrospective with prospective, another to give the prospective instance type, and a third clause to qualify this instance, often using tags or labels). However, when the query demands prospective provenance data, one needs only an extra clause (to qualify the instance), because the relationship and types will be required regardless it uses PROV-ML or not. These observations are in Table \ref{tab:q7-qualitative-analysis}. 
We verified the same behavior in the queries Q1 and Q5.

\input{our-contents/tables/q7_qual_analysis.tex}

Thus, we conclude that PROV-ML's ability to qualify specific ML data transformations and attributes using direct types eases query building, as it reduces the number of clauses required to express ML-specific concepts compared to a data representation that does not use PROV-ML. A reduction of one to three clauses per query part was observed in all queries.

\noindent
\textbf{Quantitative comparison.} We discussed in Section \ref{subsec:data_designprinciples} that certain data design decisions might accelerate queries that make use of the defined concepts, and it is known that design choices when modeling an ontology may impact the performance of queries. Also, a recent work that evaluated schema optimization to speed up queries in knowledge graphs~\cite{lei2020property}, showed that this is still a relevant topic to be investigated. Therefore, in addition to the qualitative gains discussed previously, we conduct a quantitative evaluation experiment to verify how much (if any) PROV-ML impacts query performance. 

\revisedcolor{We generate 2 synthetic datasets that mimic the real use case evaluated in Sections \ref{subsec:usecase} and \ref{subsec:performance_analysis}. With the synthetic dataset, we can control experiment variables, such as the numbers of parallel learning workflows, hyperparameters, epochs, and batches per epoch, as well as the model evaluation metrics. We can also generate one dataset that uses PROV-ML and another that does not.
If using only the real data for this experiment, it would be much harder to reproduce and control these conditions, whereas with these two synthetic datasets, it is more cost effective and feasible to switch between with and without PROV-ML, rather than having to implement, deploy, and run the real learning workflows without PROV-ML.
}
Both datasets are as follows: 8 parallel learning workflows, each with the 3 stages (training, validation, and evaluation), 300 epochs,and 200 batches per epoch (\ie{} 60,000 batches), where each batch is associated with batch losses and hyperparameters, and each epoch uses hyperparameters and generates models and model evaluations. In total, each dataset has 10,168,890 triples. 

\begin{figure}[h]
  \centering
  \includegraphics[height=3cm,keepaspectratio]{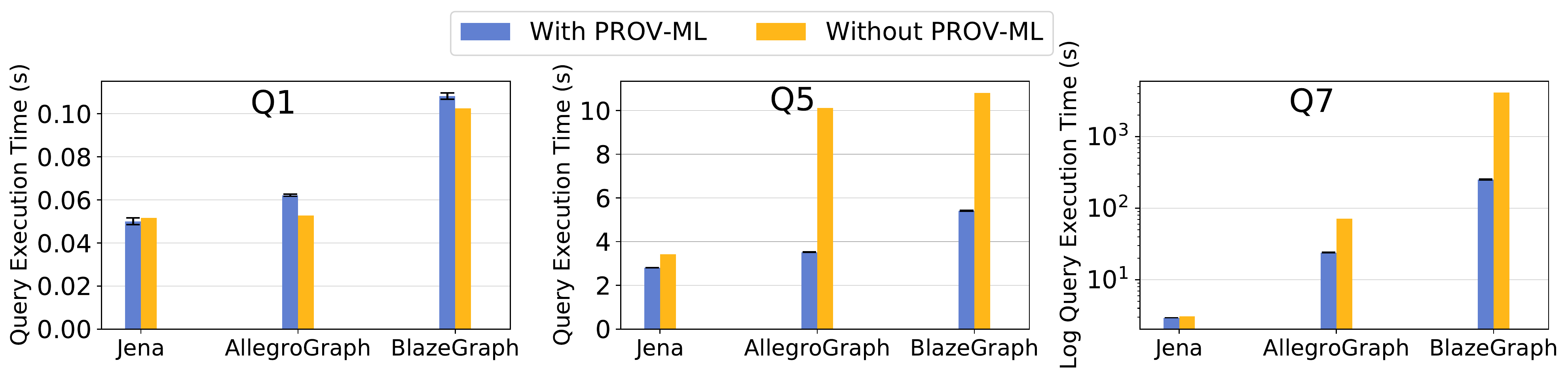}
  \caption{Execution time comparison of queries using PROV-ML vs. queries without PROV-ML. Q7 is in \textit{log} scale.}
   \label{fig:triplestoresqueries}
\end{figure}

The performance impact depends on the number of clauses to be matched in the query and on the number of triples matched by the Triple Store DBMS. However, the performance depends on the underlying DBMS that manages the \MLView{}, since the DBMS might implement efficient indexing mechanisms, parallelism techniques, or data transformation strategies. Therefore, for this experiment, we analyze the three queries (Q1, Q5, Q7).
Q1 does a simple graph traversal with simple pattern matching. Q5 does more complex graph traversals and needs to calculate aggregates (average of time difference per batch, per epoch), but for training stages only. Q7 also does complex traversals and needs to calculate aggregates (minimum batch loss per epoch), but for three stages (training, validation, and evaluation), in addition to listing hyperparameters and model performance. Since the choice of the underlying DBMS may impact the results, we analyze three different DBMSs: AllegroGraph, Blazegraph, and Jena TDB running on the same hardware and same conditions, with their default settings (no special fine-tuning are performed in any DBMS).
We analyze the query execution time, which is measured in the requesting client subtracting the timestamp obtained immediately after the response has arrived in the client from the timestamp obtained immediately before sending the request.
Results are in Figure \ref{fig:triplestoresqueries}, where we plot the medians of the query execution time over a hundred repetitions or when the confidence interval of the medians was below 5\%.
The numeric values reported in-text also refer to the median of the repetitions,
we do not remove outliers, the height of the confidence intervals are the error bars, and Q7 results are in \textit{log} scale. 

\input{our-contents/algorithms/q7_excerpts.tex}

The results show that for Q1, PROV-ML does not significantly impact query performance since the queries using PROV-ML are only {1.17x and 1.05x} slower in AllegroGraph and BlazeGraph, respectively; and the difference for Jena is within the error bars, \ie{} no statistically significant difference. 
However, Q1 is only the simplest query, with trivial graph traversals and little use of PROV-ML specific concepts, and with query times up to a hundred milliseconds (very fast queries). 
The DBMSs are likely spending more time doing data transfers than actually computing the query, which would explain the higher error bars for Q1 than for Q5 and Q7, for which the error bars are so small (\textless 1\%) that can barely be seen in the plot.
Nevertheless, Q5 and Q7 show a different behavior, particularly for the DBMSs AllegroGraph and BlazeGraph. 
In Q5, the queries with PROV-ML are {1.22x, 2.9x, and 2x} faster in Jena, AllegroGraph, and BlazeGraph, respectively. 
In Q7, they are 3x faster in AllegroGraph and 16.6x (an order of magnitude) faster in BlazeGraph, whereas in Jena, the results are nearly the same. 
Since Q5 and Q7 need to aggregate over 60,000 batches per learning stage, and Q7 has over three learning stages, these queries are considerably more complex than Q1, make much heavier use of PROV-ML specific concepts, and require complex operations in the DBMS. We analyze the system statistics while these two queries execute, and we observe full CPU utilization.
With these queries, we can see that those clauses that are reduced because of PROV-ML types make a difference in the query performance, with significant gains in AllegroGraph and BlazeGraph. In Jena, there were gains (more clearly seen in Q5), but they were not as relevant as in the other two DBMSs.
Therefore, we conclude that PROV-ML not only has qualitative advantages, such as expressiveness for ML-specific concepts and reduced complexity to write queries, but it can also help the performance of the queries, as it shows acceleration in almost all cases, particularly an order of magnitude faster for complex queries in certain workloads.

\ourvspace{}

\subsection{\revisedcolor{Utilization and Customizing the Approach to other Applications and Domains}}
\label{subsec:customization}

\revisedcolor{
Our approach is driven by queries, meaning that depending on the classes of queries (Sec. \ref{subsec:provanalysis}) the users involved in the lifecycle of the ML models
want to execute over the \MLView{} (\refDDP{dp:mlview}), they will know what to capture in their workflows. Then, they can declaratively design each workflow in the lifecycle by specifying the data transformations they want to make explicit, along with the input and output data of each transformation. In our current implementation, this specification is done using configuration files. 
To enable runtime data capture and integration in their workflows, users
import a lightweight library (PLLib) and add the data capture calls into the workflow codes (see \refSDP{dp:portable-capture}, \refSDP{dp:wf-design-hooks}). The PLLib is publicly available on GitHub \cite{provlake_github}, with examples for data transformation instrumentation.
During the specification, users also inform which data transformations are for training, validation, and evaluation, and which attributes of the transformation are hyperparameters and evaluation metrics. Such specification is used by ProvManager to create the nodes, their properties, and their relationships in the knowledge graph \MLView{}  using PROV-ML (\refDDP{dp:context-awareness}, \refDDP{dp:provschema}).
Also, in this specification, users inform (and add metadata about) the data stores being used by the workflows and the execution environments. 
Users can always revisit the set of queries they want to execute to either improve the queries (e.g., adding new fields in the projection or new filters) or add new queries, then they revisit the workflows specification and check if the hooks they add will capture the data they need for the queries. 

In addition, as we observe in Section \ref{subsec:usecase}, by 
leveraging a domain-specific knowledge graph (\refDDP{dp:context-awareness}), our approach can provide 
queries with rich semantics about the domain data.
To enable this, during the specification phase, 
users inform which data transformations and attributes are known concepts in the domain (\ie{} have been previously designed in a domain ontology), so that during the capture, ProvLake will create the relationships between the captured provenance data with the domain-specific concepts in the knowledge graph. 
This is an optional step, although it improves the overall users' understanding of the data flowing in the lifecycle, as the users typically are familiarized with the domain. 
For each application and domain, there is a specific ontology and knowledge graph with concepts and instances that make sense to that application or domain. The process of building a domain-specific ontology is out of the scope of this work. Our approach focuses on creating the relationships, in the \MLView{}, between the heterogeneous workflow data and the domain-specific knowledge graph.
}

\ourvspace{}

\subsection{Lessons Learned}
\label{sec:lessons-learned}

We draw a set of lessons learned after the practical experience of implementing the data and system design decisions to support the lifecycle in a real deployment in an O\&G industry case that uses heterogeneous environments, \ie{} a Kubernetes cluster and a large HPC cluster with CPUs and GPUs. The key findings for the success of the experiments are the following:
\begin{enumerate}[label=(\roman*),wide,labelindent=3mm,nolistsep] 
        \item \textbf{Characterizing the lifecycle and identifying the main classes for data analysis using provenance allowed the understanding of the different needs in scientific ML} (Sec. \ref{sec:ml_lifecycle}).
    Particularly, it helped to understand the different personas driving the provenance capture to answer key online and offline, intra- and inter-training provenance queries. The queries were capable of analyzing ML data, domain-specific data, and execution data, throughout the data curation, data preparation, and learning phases of the lifecycle in an integrated way. We observed that the data curation phase is the most complex. One needs to address it carefully to take advantage of domain-specific knowledge, which highly benefits trainers in the learning phase.

    \item \textbf{Employing provenance tracking and a data representation that allows data integration of multiple workflows helped to address the highly heterogeneous nature of the lifecycle.}
    To accomplish such integration, it was key to promote a holistic view of the lifecycle, end-to-end, which we called \MLView{}, as described in the data design decision \refDDP{dp:mlview}, which enabled the comprehensive data analyses (\eg{} \queries{}), thus supporting the lifecycle, which is our main motivation.
    Due to the highly heterogeneous nature, the context-awareness using domain-specific and ML data and knowledge materialized in a knowledge graph leveraging provenance-based relationships (\refDDP{dp:context-awareness}) enabled tracking, persisting, and querying interconnections between heterogeneous data with details
    about localization and data access.
     Furthermore, it enabled queries with rich semantics about the application domain and ML, exploring new data relationships that would not be possible without such context awareness. 
   
    \item \textbf{Designing a conceptual data schema focused on the key concepts enabled the design and implementation of the system facilitating query building, and the ML-specialized schema modeling}.
    The key concepts are described in \refDDP{dp:conceptual-model}.
    It enabled query acceleration and facilitated query building for queries that make heavy use of ML-specific concepts compared with a schema that does not have such specializations.
    Yet, the focused schema was the basis for PROV-ML (Sec. \ref{sec:provml}), which served as the underlying schema for the provenance system.
    PROV-ML combines the provenance of data lakes to address integration, embracing the heterogeneity nature, with concepts for ML (\refDDP{dp:provschema}).
    PROV-ML leverages W3C contributions for provenance, W3C PROV \cite{W3CPROV}, and for ML, ML Schema~\cite{W3CML}. We hope other systems with similar purposes can adopt such representation.
    
    \item \textbf{The system design decisions enabled data capture and integration in a highly heterogeneous and distributed setting adding negligible overhead.} Particularly, \refSDP{dp:portable-capture} and \refSDP{dp:microservices} provided the portability and flexibility needed in such deployments. The scalable strategies (\refSDP{dp:scalable-capture}) allowed the system to do this while incurring low overhead as well, even in HPC workloads.
\end{enumerate}

%% file: our-contents/tables/q7_qual_analysis.tex
\begin{table}[ht]
\ourvspace{}
\centering
\footnotesize
\caption{Qualitative comparison of Q7 in terms of number of clauses with and without PROV-ML.}
\label{tab:q7-qualitative-analysis}
\topsepremove
\begin{tabular}{lcc}
\rowcolor{table_header}
\multicolumn{1}{l}{\textbf{Q7 Query part}} & \textbf{\#Clauses w/ PROV-ML} & \textbf{\#Clauses w/o PROV-ML} \\ \toprule
\rowcolor{table_row_even}
Training stage                   & 1                             & 4                              \\ 
\rowcolor{table_row_odd}
Epoch iteration                  & 2                             & 5                              \\ 
\rowcolor{table_row_even}
Model hyperparameters            & 6                             & 7                              \\ 
\rowcolor{table_row_odd}
Model                            & 2                             & 5                              \\ 
\rowcolor{table_row_even}
Model evaluation                 & 3                             & 6                              \\  \arrayrulecolor{table_row_odd}\hline
\end{tabular}
\ourvspace{}
\end{table}

%% file: our-contents/algorithms/q7_excerpts.tex
\noindent
\begin{minipage}{.45\textwidth}
\begin{lstlisting}[
    language=sql,
    caption={ Excerpt of Q7 with PROV-ML. },
    label={listing:q7-provml},
    keywords={a, prov, provml, provlake}
]
-- Training stage
?training a provml:TrainingExecution.
-- Epoch iteration (Training section)
?epoch_exec_training 
    prov:wasInformedBy ?training;
    a provml:TrainingSectionExecution.
-- Model hyperparameters
?epoch_training_hyperparam
    prov:wasGeneratedBy ?epoch_exec_training;
    a provml:ModelHyperparameterValue;
    prov:value ?epoch_training_hyperparam_v;
    prov:wasDerivedFrom ?epoch_training_hpram_psp.
?epoch_training_hpram_psp 
    a provml:LearningHyperparameterSetting;
    rdfs:label ?epoch_training_hyperparam_name.
-- Model
?model_training 
    prov:wasGeneratedBy ?epoch_exec_training;
    a provml:Model.
-- Model evaluation
?model_training_eval 
    prov:wasGeneratedBy ?epoch_exec_training;
    a provml:ModelEvaluation;
    prov:value ?model_training_eval_value.

\end{lstlisting}

\end{minipage}
\begin{minipage}{.03\textwidth}
\
\end{minipage}
\begin{minipage}{.45\textwidth}
\begin{lstlisting}[
    language=sql,
    caption={ The same excerpt of Q7, without PROV-ML. },
    label={listing:q7-noprovml},
    keywords={a, prov, provml, provlake}
]
-- Training stage
?training a
    provlake:DataTransformationExecution;
    prov:wasInfluencedBy ?training_prosp;
?training_prosp
    a provlake:DataTransformation;
    provlake:tag "Training".
-- Epoch iteration (Training Section)
?epoch_exec_training 
    prov:wasInformedBy ?training;
    a provlake:DataTransformationExecution;
    prov:wasInfluencedBy epoch_exec_training_psp.
?epoch_exec_training_psp
    a provlake:DataTransformation;
    rdfs:label "Epoch Execution".
-- Model hyperparameters
?epoch_training_hyperparam
    prov:wasGeneratedBy ?epoch_exec_training;
    a provlake:AttributeValue;
    prov:value ?epoch_training_hyperparam_v;
    prov:wasDerivedFrom ?epoch_training_hpram_psp.
?epoch_training_hpram_psp 
    a provlake:Attribute;
    provlake:tag "Hyperparameter";
    rdfs:label ?epoch_training_hyperparam_name.
-- Model    
?model_training 
    a provlake:AttributeValue.
    prov:wasGeneratedBy ?epoch_exec_training;
    prov:wasDerivedFrom ?model_training_prosp.
?model_training_prosp 
    a provlake:Attribute;
    provlake:tag "Model".
-- Model evaluation
?model_training_eval 
    prov:wasGeneratedBy ?epoch_exec_training;
    a provlake:AttributeValue;
    prov:value ?model_training_eval_value;
    prov:wasDerivedFrom ?model_training_eval_psp.
?model_training_eval_psp 
    a provlake:AttributeValue;
    provlake:tag "Model Evaluation".

\end{lstlisting}  
\end{minipage}

%% file: our-contents/sections/related_work.tex
\ourvspace{}
\section{RELATED WORK} \label{sec:related_work}

The interest in workflow provenance management has increased in the recent years, driven by a major effort by the provenance community \cite{thavasimani2018exploiting,missier2020abstracting,lourenco_sigmod2020,mehta_2019,garijo_integrating_2018,namaki_vamsa_2020,sikos_provenance_aware_2020,alessandro_escience2019,guedes_capturing_2020,goble_fair_2019,magnoni_dare_2019,tale_2020,mcphillips_yesworkflow_2015,pimentel_noworkflow_2017,rupprecht13improving}
, particularly to explore possibilities of optimizing workflows with the data captured by provenance tools and as a response to the urgent need for reproducible science, which is critical in scientific ML \cite{factsheets_ai_ibm_2019}.
To exemplify, Thavasimani~\etal{} \cite{thavasimani2016facilitating} investigate provenance traces recorded during workflow executions to observe differences in results with minor workflow configuration differences. 
Other works have advanced provenance tracking techniques on heterogeneous data, stores, and environments \cite{komadu_escience2016,hu2019efficient, polyflow_2019,datalake_vldb19} and others have explored the intersection of provenance and blockchain \cite{liang2017provchain,prov_blockchain_vldb_2019}. 
 
On the intersection between ML and provenance, other works have explored provenance to support ML workflows \cite{miao_towards_2017,miao_sigmod2019,kumar_model_2016,zaharia_accelerating_2018,debora_sbbd_2019,souza_dlsteer_2018} and
  Deelman \etal{} \cite{RoleMLinWorkflows} characterized provenance analysis to leverage ML in support of scientific workflows.  
 On reproducible ML models, another aspect that has been explored is the use of provenance as an essential tool to help create explainable artificial intelligence~\cite{lucero2018exploring,factsheets_ai_ibm_2019}. In addition, some works addressed the gap between the experiments of an ML workflow execution and a standard representation to provide reproducible experiments~\cite{esteves2015, publio2018, W3CML}. Esteves~\etal{}~\cite{esteves2015} provide a machine-readable vocabulary and a common schema for reproducibility of ML experiments in various frameworks and workflow systems. Publio~\etal{}~\cite{publio2018} present a new ML data representation based on MEX vocabulary \cite{esteves2015} to improve processes on ML workflows, despite not having a clear separation between prospective and retrospective provenance. Samuel~\cite{samuel_machine_2020} propose ProvBook, for reproducibility of ML experiments using Jupyter notebooks applying FAIR data principles.
 Moreno \etal{} proposed MLWfM~\cite{moreno_managing_2019} to provide data concepts for ML and domain-specific awareness, but without provenance concepts and a data representation. 
Brandao \etal{} \cite{brandao2020knowledge} proposed a knowledge-based workflow management approach aiming at broadening user collaboration over ML experiments. It provided a semantic structure for computational workflows allowing rich querying at different levels of provenance.

These works are important building blocks to support the \MLCycle{} using provenance management techniques. 
Nevertheless, they still lack a holistic view capable of comprehensively integrating the data in the whole lifecycle, end-to-end, from raw domain data to learned models.
Without such a holistic view, the ML-specific concepts cannot integrate with the specific concepts about the scientific domain, jeopardizing the comprehensive end-to-end analyses that require richer semantics about the domain integrated with rich semantics about ML.

%% file: our-contents/sections/conclusion.tex
\ourvspace{}
\section{CONCLUSIONS} \label{sec:conclusion}

In this work, 
we aimed at enabling scientists and engineers to perform comprehensive data analyses in the \MLCycle{}. 
We proposed workflow provenance techniques to address the problem of dealing, in an integrated and comprehensive way, with the high heterogeneity of different contexts 
(\eg{} data, software, environments, persona) involved in the lifecycle, to enable such analyses. 
We proposed modeling the workflows in all phases of this lifecycle as multiple interconnected workflows. A holistic view of the data processed in these workflows should be built as the workflows execute. In this way, the collaborating teams can use it as their primary source of data analyses that integrate that from raw data to learned ML models.
We called it as \textit{Provenance-based Holistic Data View of the \MLCycleCammel{}} (\MLView{}).
It is materialized as a knowledge graph with provenance-based relationships. It is aware of the contexts of the data transformations in the workflows, their (hyper)parameterizations and model metrics, which computational environments they run and data stores they use, the involved personas, and how they interact with the workflows.

To be able to build this view, aware of these many 
dimensions of heterogeneity, we first characterized the lifecycle and proposed a taxonomy for the classes of data analyses (\eg{} data, execution timing, and training timing).
Then, we proposed design principles for the effective and efficient management of provenance data from these workflows.
From this understanding and design principles, we derived the PROV-ML data representation, promoting such a holistic view of data in workflows in the lifecycle, which is the first one to the best of our knowledge.

We also proposed system design principles and a reference system architecture to provide the view with efficient provenance capture adding significant data capture overhead (\textless1\%). It allows for portable and flexible deployments required because of the heterogeneous executions. We obtained these results after implementing the design principles and the PROV-ML in the ProvLake system and deploying it in a real case in the O\&G industry.
Altogether, based on our studies, our major finding is that the design principles enabled comprehensive queries, with rich semantics about the domain and ML, by exploring the data view while maintaining high scalability even in HPC workloads. 
\revisedcolor{Finally, ongoing work is towards applying our approach in the finance industry, to integrate workflows in cloud and HPC clusters to train ML models for credit risk assessment. The preliminary results indicate that our approach can be customized with the effort described in Section \ref{subsec:customization}. }

\ourvspace{}

%% file: our-contents/sections/appendices.tex
\newpage
\appendix




\section{Details on the data captured by ProvLake using PROV-ML in the use case.}
\input{our-contents/tables/prov-use-case.tex}

%% file: our-contents/tables/prov-use-case.tex
\begin{table}[ht]
\footnotesize
\caption{Details about the captured data in the use case.}
\label{tab:use-case}
\begin{tabular}{>{\arraybackslash}m{0.14\textwidth}m{0.35\textwidth}m{0.35\textwidth}m{0.08\textwidth}}
\hline
\rowcolor{table_header}
\textbf{Data structure name} & \textbf{Description} & \textbf{Data Characteristics and Size} & \textbf{Data Store} \\
\rowcolor{table_row_even}
Geoscientist's Annotations &
Observations they do about the seismic dataset, such as its geographic global coordinates and characteristics about the subsurface terrain this seismic acqusition was obtained. Also, they relate the seismic datasets with other artifacts of interest, such as well logs and geological basins.
&
Semi-structured textual files
&
Textual documents in the file system
\\
\rowcolor{table_row_odd}
Structured domain knowledge &
Domain-specific information parsed from unstructured and semi-structured documents and represented as structured facts in domain ontologies. Entities in such ontologies may represent taxonomy, rules and assertions for a given domain. &
Stored as domain-specific knowledge graphs in a Knowledge Base, typically managed by a Triple Store &
Triple store \\
\rowcolor{table_row_even}
Geological Labeled Data &
Tabular text files, where each line contains x, y positions (floar32, float32) on Earth surface, and depth (float32) that can be in distance or time. &
$N_x \cdot N_y \cdot N_h \cdot 4$ bytes, where $N_x$ and $N_y$ are the number of points in $x$ and $y$ directions, respectively. $N_h$ is the number of annotated horizons. 
&
File system
\\
\rowcolor{table_row_odd}
Post-stacked SEGY file &
A binary file containing $N_x \times N_y$ stacked traces of one particular seismic attribute, e.g. amplitude, coherence, frequency, phase. The file also includes a main header and several trace headers. &
$H_{main} + N_x \cdot N_y \cdot ( H_{trace} \cdot T_{size})$, where the main header ($H_{main}$) takes approximately 10KB; the trace header uses 240 bytes; and each trace contains one float32 value for each point in depth. For example, if the seismic is a volume $1000 \times 2000 \times 3000$, besides the headers, it will contain $2 \times 10^{6}$ traces each of which comprising $3000$ float32 values. &
File system \\
\rowcolor{table_row_even}
Curated and annotated seismic data &
Merged expert annotations and the SEGY raw file. It comprises the structured knowledge about the geological data and also cube geometry, such as inline and crossline ranges, resolution, depth range, and unit. The expert informs which parts of the input file are suitable or not for the task. Finally, it may contain legal and access information. With this data, it is possible to set next phase hyperparameters. &
Stored as structured data in a combination of Doc. DBMS and Knowledge Bases with references to the Doc. DBMS. The Doc. DBMS has hundreds of gigabytes and the Knowledge Base has hundreds of megabytes.
 &
Triple Store and Doc. DBMS \\
\rowcolor{table_row_odd}
Training, validation, and evaluation data sets &
Binary files stored in HDF5 or using Google's Protocol buffer serialization for a good balance between portability and speed. &
These files vary depending on the configuration selected for data preparation workflows. From our experimental observations, it takes about 10\% or less of the input SEGY file. However, because workflows create data sets by experiment configuration, it is possible to end up with a total data set storage multiple times bigger than the original raw file. &
Cloud Object Store and 
File system \\
\rowcolor{table_row_even}
Learned models &
Mix of binary and configuration files depending on the engine used to run the learning phase (PyTorch, Tensorflow, Scikit-learn, etc.). &
The engine used to run defines trained models' type and size. Since we used Tensorflow backend in our experiments, we store our trained models using Tensorflow's tools, where each experiment produces configuration and binary files. The first one stores the model structure and other training parameters, and the second one stores the model's state. Although model size can vary from a few MB to several GB, our models used approximately 50MB per state in our experiments. Notice that one state is just one snapshot of one step during training, so if depending on the configuration settings, it is possible to have several saved states, the 50MB may turn GB very quickly. &
File system \\
\arrayrulecolor{table_row_odd}\hline
\end{tabular}
\end{table}